\documentclass[letterpaper,twocolumn,10pt]{article}

\usepackage{tikz}
\usepackage{amsmath}
\usepackage{framed}
\usepackage{graphicx}
\usepackage{xurl}
\usepackage{multirow}
\usepackage{hyperref}
\usepackage{subcaption}
\usepackage{filecontents}
\usepackage{float}
\usepackage{enumitem}
\usepackage[font=small,skip=0pt]{caption}
\setlist{nosep}
\restylefloat{table}

\begin{document}

%don't want date printed
\date{}

\title{\Large \bf Designing and Testing a Mobile Application for Collecting WhatsApp Chat Data While Preserving Privacy}

\def\plainauthor{Anonymous Author(s)}

\author{
{\rm Brennan Schaffner}\\
University of Chicago
\and
{\rm Archie Brohn}\\
University of Chicago
\and
{\rm Jason Chee}\\
University of Chicago
\and
{\rm K.J. Feng}\\
Princeton University
\and
{\rm Marshini Chetty}\\
University of Chicago
} % end author

\maketitle

\begin{abstract}
   It is common practice for researchers to join public WhatsApp chats and scrape their contents for analysis. However, research shows collecting data this way contradicts user expectations and preferences, even if the data is effectively public. To overcome these issues, we outline design considerations for collecting WhatsApp chat data with improved user privacy by heightening user control and oversight of data collection and taking care to minimize the data researchers collect and process off a user's device. We refer to these design principles as User-Centered Data Sharing (UCDS). To evaluate our UCDS principles, we implemented a mobile application representing one possible instance of these improved data collection techniques and evaluated the viability of using the app to collect WhatsApp chat data. Second, we surveyed WhatsApp users to gather user perceptions on common existing WhatsApp data collection methods as well as UCDS methods. Our results show that we were able to glean similar informative insights into WhatsApp chats using UCDS principles in our prototype app to common, less privacy-preserving methods. Our survey showed that methods following the UCDS principles are preferred by users because they offered users more control over the data collection process. Future user studies could further expand upon UCDS principles to overcome complications of researcher-to-group communication in research on WhatsApp chats and evaluate these principles in other data sharing contexts.
\end{abstract}

\section{Introduction}
\label{introduction}
WhatsApp, the most popular mobile messenger application worldwide, has two billion global active users~\cite{noauthor_two_2020}. Researchers have extensively studied WhatsApp, often focusing on aspects such as the structure of sharing networks~\cite{resende_misinformation_2019} or the spread of election misinformation~\cite{garimella_images_2020, machado_study_2019, melo_whatsapp_2019, moreno_whatsapp_2017, narayanan_news_2019, reis_dataset_2020}. To study WhatsApp data, it has been common practice for researchers to join public WhatsApp groups and scrape their contents for analysis. However, recent work---on platforms other than WhatsApp---has shown that collecting public data in this way may contradict the expectations and preferences of users or violate the integrity of the researcher--public relationship, even if the content is effectively public~\cite{fiesler_participant_2018,zimmer_but_2010, cooper_ethics_2020, hagen_all_2020, mancosu_what_2020}.
The notion of `public' data is further complicated by the fact that WhatsApp is end-to-end encrypted by design, so users on the platform may expect a degree of privacy more so than on the open Internet. In fact, contemporary privacy frameworks like contextual integrity (CI)~\cite{nissenbaum2004privacy, nissenbaum2019contextual} highlight the importance of upholding contextual norms of information transfer, meaning scraping data from WhatsApp chats may violate user privacy since it violates norms and expectations around data sharing in a chat.
Therefore, we as a research community should consider revisiting best practices for collecting and analyzing WhatsApp data that may be more in line with users' expectations and wishes or risk eroding the public's trust. 

The benefits of studying WhatsApp data are clear, especially in the field of misinformation. Journalists, fact-checking organizations, and academic researchers have highlighted that WhatsApp is an effective pipeline for misinformation \cite{tijani_how_2020, purohit_misinformation_2020, hitchen_whatsapp_2019, shahid_it_2022, feng_investigating_2022}. Misinformation spreading on WhatsApp has resulted in severe consequences, such as when fabricated content sparked dozens of killings~\cite{kaur_information_2018, bengali_how_2019}, or when viral misleading content influenced a presidential election~\cite{avelar_whatsapp_2019}. 
Additionally, the platform serves as a primary means of communication on the Internet for many.
At least 95\% of the populations of Nigeria, South Africa, and Kenya use WhatsApp monthly~\cite{statista_mobile_nodate}.
Thus, WhatsApp's two billion global active users~\cite{noauthor_two_2020} are at risk of consuming misinformation on the same platform they use as a primary means of Internet communication. 
Understanding misinformation on WhatsApp has direct importance for suppressing its spread and mitigating its impact. These works that study the spread and consequences of WhatsApp's viral misleading content highlight the need for continued research using WhatsApp data.

Yet, scraping and analyzing publicly accessible data has caused ethical concerns, e.g., a majority of Twitter users report not knowing that researchers use tweets and believing that researchers should have to obtain permission to do so~\cite{fiesler_participant_2018}. Investigating more privacy-preserving techniques for collecting WhatsApp chat data would therefore benefit the research community. Thus, we posed the following research questions: 
\begin{enumerate}
    \item \textbf{RQ1}: How can we study WhatsApp chats in a way that gives users more control over what data researchers collect? 
    \item \textbf{RQ2}: What can we learn about WhatsApp chats using such a technique? 
    \item \textbf{RQ3}: What do WhatsApp users think about different methods of collecting their chat data?
\end{enumerate}

To answer these questions, we first outline four principles for collecting WhatsApp chat data that give users more control and oversight of the data collection and ensure researchers minimize the data they collect and process off a user's device. 
We refer to these principles as the User-Centered Data Sharing (UCDS) principles.
Briefly, the principles are: \textit{Constrained Data Collection} restricts data collection to strictly necessary metadata; \textit{Local Extraction and Processing} minimizes the data researchers collect and process off a user’s device; \textit{User Involvement} gives participants control and oversight of their data; and \textit{Transparency} equips users with information needed to make informed decisions about sharing their data. 
We intentionally use `data sharing' instead of `data collection' to indicate the mutual relationship between researchers and users unlike the single-sided notion of data collection that asymmetrically empowers researchers. 
After outlining the UCDS principles, we then created a mobile application for iOS called URL-EXTRACTOR-APP following UCDS principles to evaluate whether UCDS still allows researchers to glean useful insights from WhatsApp chats. Specifically, the app allows WhatsApp users to voluntarily share limited content from their WhatsApp chats and have control over the data being collected. We then conducted a user study to evaluate the feasibility of using URL-EXTRACTOR-APP to collect private chat data with 10 
participants in the United States (US). We focused on private WhatsApp chats for the feasibility deployment---as opposed to larger public ones\footnote{We use the same distinction of \textit{private} and \textit{public} chats as Feng et al.~\cite{feng_investigating_2022}, where chats joined by users via direct communication with a user’s existing mobile contacts are \textit{private} WhatsApp chats and chats where invite links are posted online so that anyone can join are \textit{public} chats.}---since they potentially represent a stronger case of privacy requirements as they are made up of more intimate relationships (friends and family) and fewer members than the large public chats \cite{seufert, feng_investigating_2022}. Yet messages from these chats can still propagate widely owing to WhatsApp message forwarding \cite{melo-forwarding}. Most crucially, private WhatsApp chats are largely understudied yet make up a majority of WhatsApp chats~\cite{melo-forwarding}. From the initial deployment of URL-EXTRACTOR-APP, we found that using the mobile application that ran locally on users' phones to extract metadata from WhatsApp chats was a feasible technique for gathering WhatsApp chat data and making discoveries about WhatsApp private chats' contents while following UCDS principles.
From the collected data using UCDS, we were able to discern that a single chat member consistently shared a majority of links in each chat, users shared links relatively infrequently during the time period of chats in the dataset, and shared links covered a variety of topics with \textit{YouTube} and \textit{Google} links being the most common. Although our results are not generalizable owing to a small sample, they demonstrate that our set of UCDS best practices for gathering private WhatsApp data remain effective and that we can glean valuable insights about private WhatsApp chats without explicitly collecting all the chats' available contents, such as the text messages or images sent in the chat. 

Second, we surveyed 334 WhatsApp users to gauge their perceptions of various approaches to WhatsApp data collection including UCDS principles.
The survey results indicate that following UCDS principles improves upon common WhatsApp data collection methods. Specifically, common existing data collection
methods that target and scrape public WhatsApp chats violate user expectations and miss a majority of the WhatsApp
messaging ecosystem by focusing on public chats only. 
Participants preferred how methods aligning with UCDS principles limited data collection to metadata, sanitized and anonymized the data, and showed users what data was being collected all before it reached the researchers.
Our survey results also suggested that UCDS principles can be expanded to address the complications of researcher-to-group communication. Based on our findings, we make recommendations for improvements to best practices in data collection techniques more broadly.

Our main contributions are summarized below:
\begin{itemize}
    \item We provide evidence that new data collection methods for WhatsApp are needed for two reasons. First, corroborating prior work~\cite{seufert}, we find that the most popular usage of WhatsApp is among friends and family, suggesting that a majority of WhatsApp content is in private chats. Therefore, existing methods targeting publicly accessible chats may be missing a majority of WhatsApp content. Second, analogous to Fiesler et al.\ finding that research using public Twitter data violated user wishes~\cite{fiesler_participant_2018}, we show that WhatsApp users are not comfortable with common WhatsApp research methods, indicating the necessity of revisiting best practices.
    \item We outlined UCDS principles by building on common privacy practices from prior works for WhatsApp data collection~\cite{garimella_whatsapp_2018, javed_first_2020, machado_study_2019, resende_misinformation_2019, resende_analyzing_2019, maros_analyzing_2020, narayanan_news_2019, saha_short_2021} and adding additional ethical best practices including: restricting collection to metadata, local data extraction and processing, increased participant control over data collection, and explicit transparency in the data collection methodology. We present a research method for collecting WhatsApp chat data using a mobile app that exemplifies one possible implementation of UCDS principles. 
    \item We demonstrate the effectiveness of UCDS principles through a feasibility study of our app to collect private WhatsApp chat data while upholding participant oversight of data collection, and present insights from an exploratory dataset of private WhatsApp chat data from study participants in the US.
    \item We provide evidence that users support WhatsApp UCDS best practices for sharing their WhatsApp chats. 
\end{itemize}

\section{Background and Related Work}
\label{sec:lit_rev}

We provide an overview of the WhatsApp platform and present common approaches to WhatsApp data collection and uses of public WhatsApp chat data. 

\begin{figure*}
    \centering
    \begin{subfigure}[t]{0.24\textwidth}
        \includegraphics[width=\textwidth]{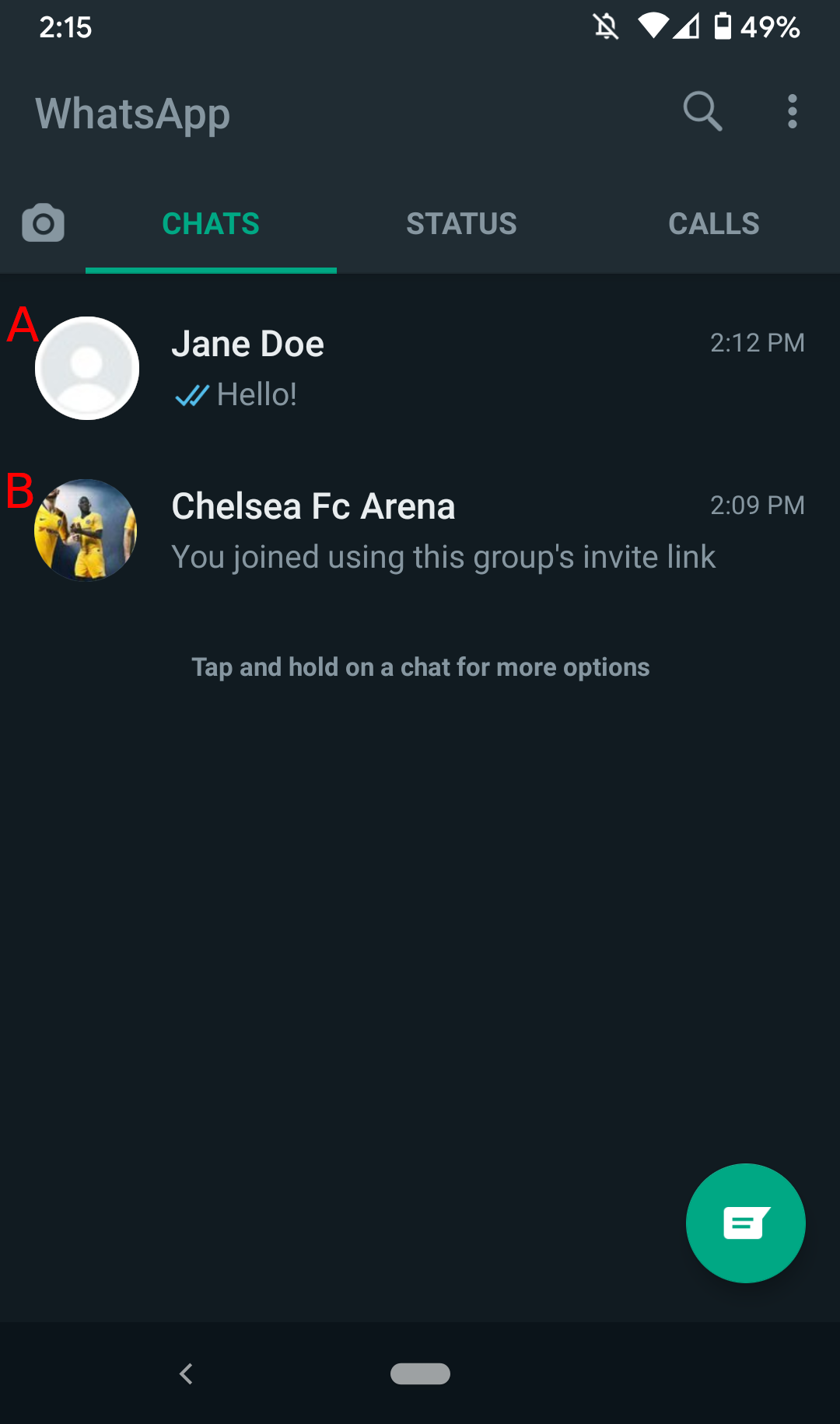}
        \caption{}
        \label{fig:chattypes}
    \end{subfigure}
    \hfill
    \begin{subfigure}[t]{0.24\textwidth}
        \includegraphics[width=\textwidth]{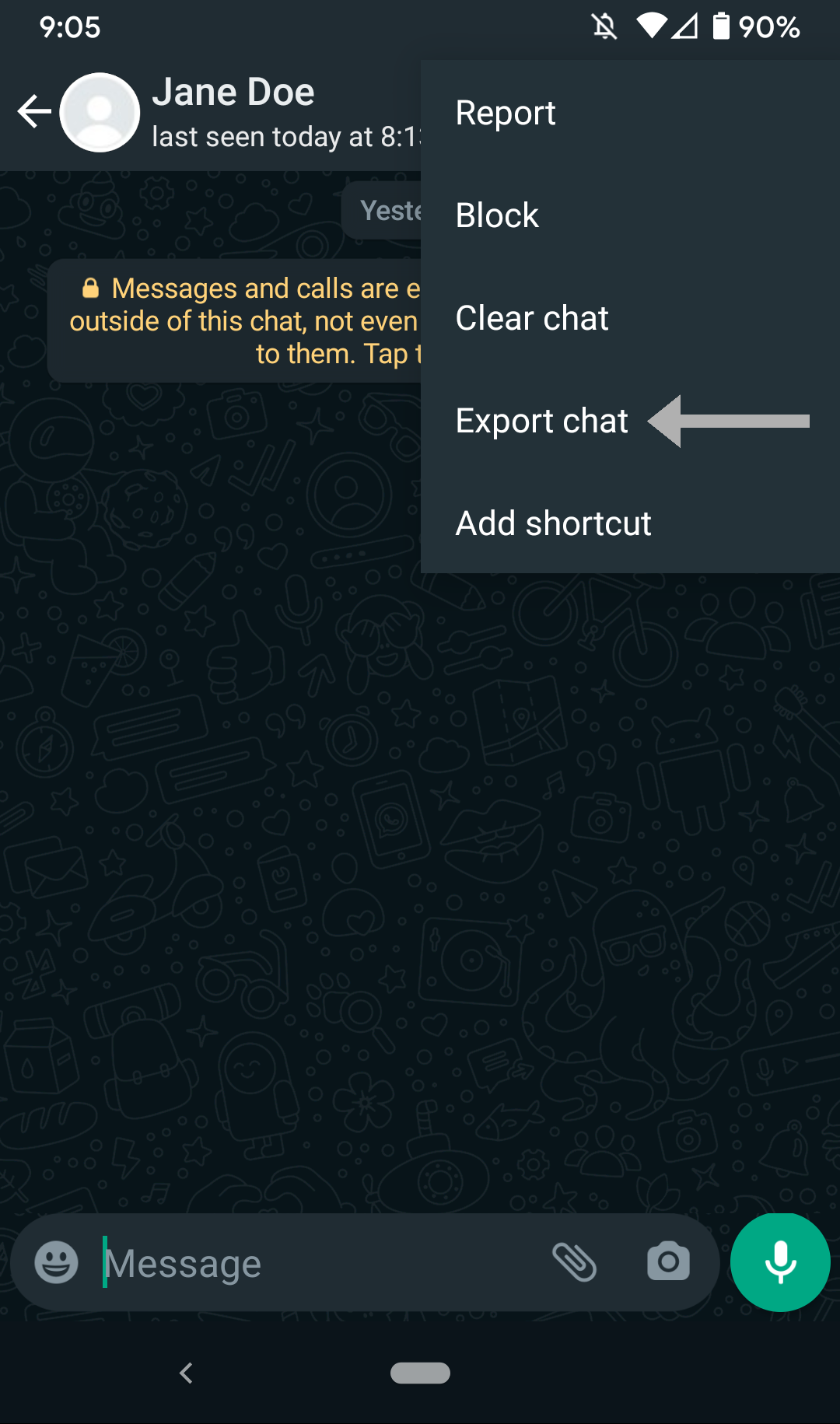}
        \caption{}
        \label{fig:chatoptions}
    \end{subfigure}
    \hfill
    \begin{subfigure}[t]{0.24\textwidth}
        \includegraphics[width=\textwidth]{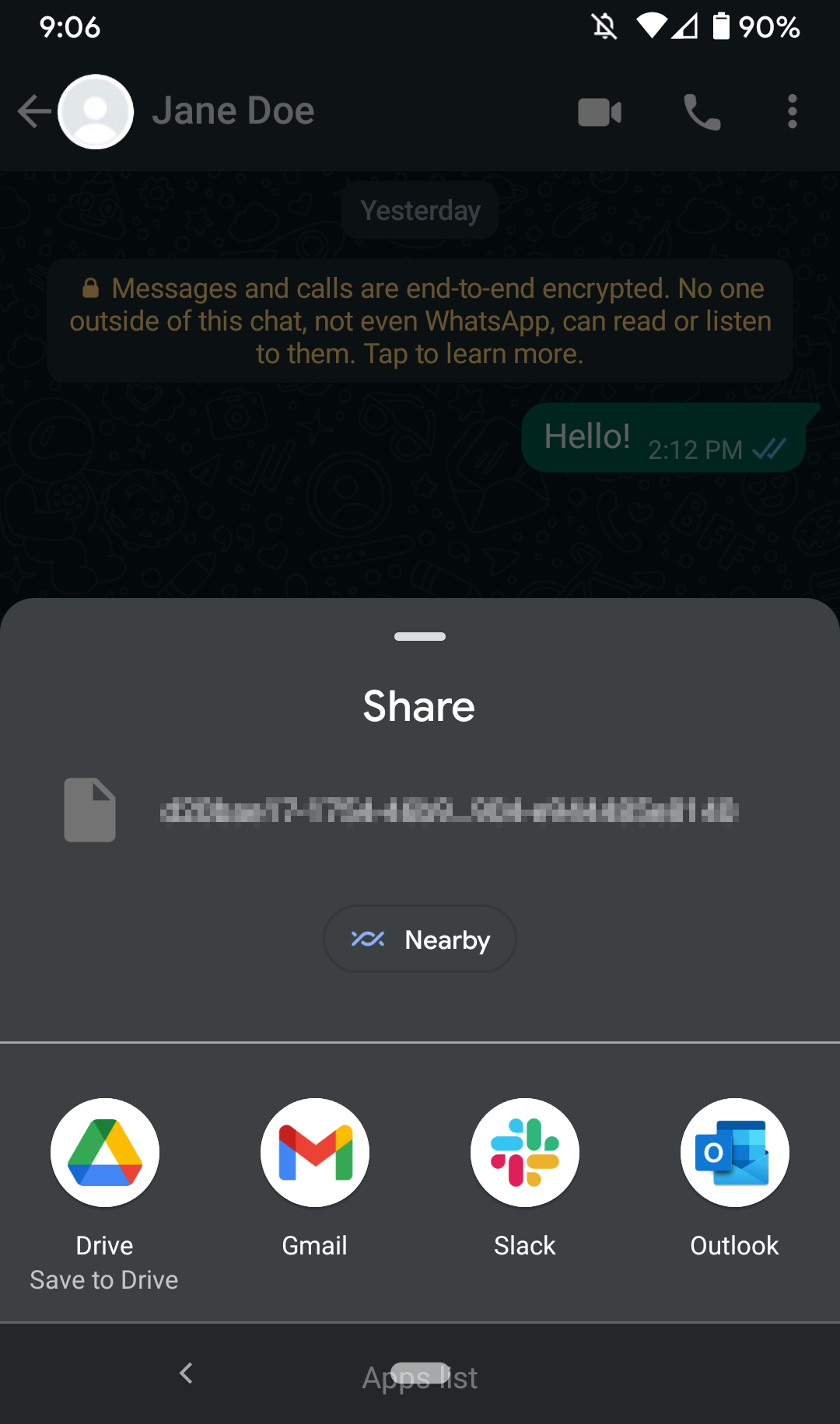}
        \caption{}
        \label{fig:chatshare}
    \end{subfigure}
    \hfill
    \begin{subfigure}[t]{0.24\textwidth}
        \includegraphics[width=\textwidth]{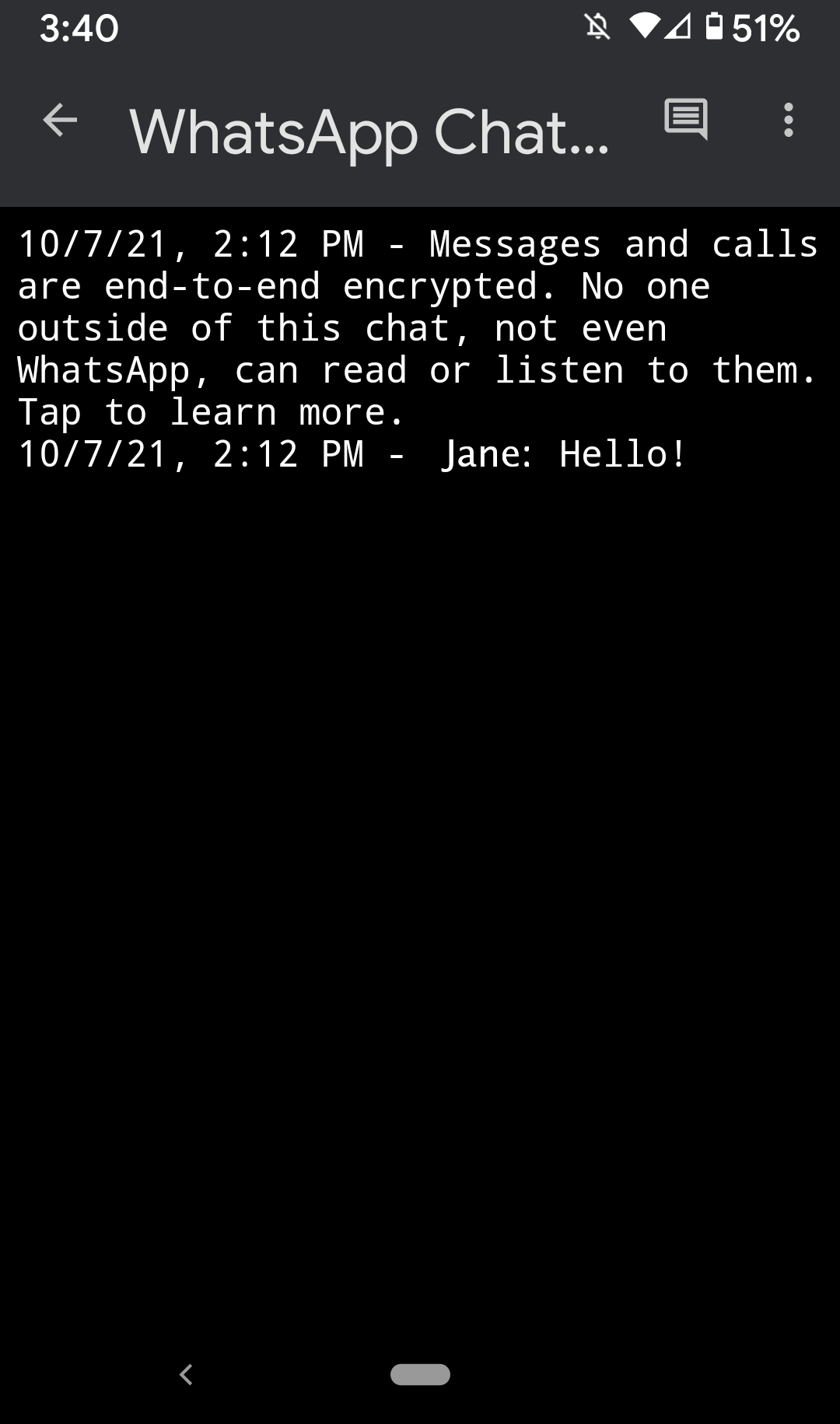}
        \caption{}
        \label{fig:textexample}
    \end{subfigure}
    \caption{The process for exporting the text file logs from a WhatsApp chat and an example of the resulting text file. WhatsApp's homepage is shown in (a), where users can select one of their chats. Then in (b) the user can select to export the chat data from the chat's options menu. Finally, the user chooses how to export the chat in (c), linking to other apps that accept text files. The resulting exported chat file is shown in (d).
    }
    \label{fig:extractprocess}
\end{figure*}

\subsection{WhatsApp: Private and Public Chats}
WhatsApp, a mobile messaging application based in the US owned by Meta, was released in 2009 (for iOS) and 2010 (for Android)\cite{iphone_whatsapp_2019, sawers_three-quarters_2015,tsotsis_whatsapp_2014}.
WhatsApp supports sharing images~\cite{olson_exclusive_2014}, audio messages~\cite{ho_voice_2013}, documents~\cite{perez_whatsapp_2016}, and videos~\cite{hatmaker_whatsapp_2021}, and WhatsApp contacts can also communicate via voice calls and video meetings~\cite{chowdhry_whatsapp_2016, vincent_whatsapp_2015}. Notably, all communication is encrypted by default~\cite{metz_forget_2016}. The platform's popularity is in part due to allowing users to send messages over a device's Internet connection so they can avoid expensive Short Message Service (SMS) fees~\cite{hoffman_how_2013}. In 2019, there were 67.1 million US WhatsApp users with projections of 85.8 million users in 2023~\cite{statista_whatsapp_nodate}.

Conversations on WhatsApp take the form of ``chats''. 
The platform allows users to participate in multiple chats simultaneously. Users can be added to group chats by a group administrator, or they can join chats via invite links obtainable from existing chats. Additionally, they can create new chats and add their WhatsApp contacts. 
Figure~\ref{fig:chattypes} shows an example private chat with a friend (A) and public chat for fans of a popular football club (B).
Websites like \texttt{WhatsApp Group Links}~\cite{whatsappgrouplinks} can host a list of ready-to-join public WhatsApp groups (up to 257 members), often centered around a common interest (e.g.,\ political opinions, shopping deals, or dating). 
Another feature provided by WhatsApp is the ability to export chat information. Due to the forthcoming relevance of this feature, we outline the process in Figure~\ref{fig:extractprocess}.

\subsection{Research Using Public WhatsApp Data}
Prior academic work has extensively collected and analyzed WhatsApp chat data.
In 2018, Garimella and Tyson explored the feasibility of collecting and using WhatsApp data~\cite{garimella_whatsapp_2018}. They present an example methodology used to analyze a random sample of 178 public group chats that they joined using online invite links and automated web scraping tools. 
Like our study, this work explored what WhatsApp group data can be collected and how it can be used. However, our work differs in that it focuses on developing methods for collecting WhatsApp chat data with additional privacy considerations.

Building on the feasibility of collecting and analyzing public WhatsApp group data, some researchers have studied the network structures of the spread of misinformation related to major social events~\cite{resende_misinformation_2019} and elections~\cite{melo_whatsapp_2019, moreno_whatsapp_2017}. Other researchers have analyzed public WhatsApp groups for specific types of harmful content such as ``fear speech''~\cite{saha_short_2021} and incitements to violence in India~\cite{arun_whatsapp_2019} or COVID-19 misinformation in Pakistan~\cite{javed_first_2020}.

Some researchers have focused on a specific type of content (images, audio, etc.) spreading amongst WhatsApp groups.
A large portion of WhatsApp studies focused on the spreading of image content~\cite{machado_study_2019, narayanan_news_2019, garimella_images_2020, reis_dataset_2020, resende_misinformation_2019}. For example, Garimella and Eckles found that 13\% of image shares contained misinformation when analyzing politically oriented public WhatsApp groups in India. Researchers have also examined text-based misinformation on WhatsApp. For example, Resende et al.\ focused on text-only messages, finding that textual misinformation spread quickly within social groups but took longer to bridge into others~\cite{resende_analyzing_2019}. Other studies analyzed the uniform resource locators (URLs)~\cite{resende_analyzing_2019, machado_study_2019, narayanan_news_2019}, audio messages~\cite{maros_analyzing_2020}, or video content~\cite{machado_study_2019} shared in WhatsApp chats. 
Some misinformation studies on WhatsApp did not directly analyze data from WhatsApp but used interviews to probe about (mis)information sharing dynamics~\cite{feng_investigating_2022, karusala_towards_2022}.
We extend prior work to explore how to collect and analyze data from private WhatsApp chats without ever directly observing actual chat messages and giving participants oversight of the data collection.

\subsection{Collecting WhatsApp Data}
\label{commonpractices}
Academics have broadly discussed the ethics of general online data collection~\cite{fiesler_participant_2018, vitak_ethics_2017, tranberg_dataethics_2018, buchanan_internet_2021, monkman_ethics_2018, vitak2016beyond, pater2022no, gilbert2021measuring} and web scraping/crawling~\cite{gold_robots_2018,  landers_primer_2016, rennie_scraping_2020, luscombe_algorithmic_2021, brewer_ethics_2021, krotov_legality_2018} on other platforms, but analogous research is lacking for WhatsApp, perhaps due to the added complexity brought by its encryption.
That is, even though WhatsApp group invite links may be public, their message content is still encrypted and may not be considered public to the same extent as unencrypted public content on other platforms. 
This is demonstrated by the fact that researchers must join public WhatsApp groups to access and scrape their contents. 
WhatsApp researchers have weighed privacy considerations of their data collection in a variety of ways. We identified three common practices that researchers use to protect the privacy and wishes of their participants. First, some researchers anonymize the personally identifiable information (PII) of the WhatsApp chat members by creating cross references between unique IDs and their identifiable information from their WhatsApp accounts (e.g.,\ names and phone numbers)~\cite{garimella_whatsapp_2018, javed_first_2020, machado_study_2019, resende_misinformation_2019, resende_analyzing_2019, maros_analyzing_2020, reis_dataset_2020, narayanan_news_2019, saha_short_2021, schwind2018whatsanalyzer}. This protects users because researchers can delete phone numbers from their datasets before analyzing the collected data. 
Second, some researchers constrain the data collected from public WhatsApp chats limiting the collection of inessential data, which may include PII. For example, researchers may only collect a certain type of message such as images or news stories~\cite{machado_study_2019, narayanan_news_2019}.
Third, some researchers announce their presence in the chats from which they plan to begin scraping data by posting a message with their intentions~\cite{machado_study_2019, narayanan_news_2019}. These disclosure messages inform the chat members that they can voice their concerns or request to opt-out of the study. 
In practice, researchers have used all, some, or none~\cite{garimella_images_2020, melo_whatsapp_2019, moreno_whatsapp_2017, shah_tools_2019, karusala_towards_2022} of these methods in their WhatsApp studies. 

While these three common practices strongly work in favor of participant privacy, there are still some ethical concerns with the way the data is being collected. The framework provided by these practices defaults WhatsApp users into a position of trusting researchers. Even if researchers announce their presence and tell users they can opt-out, how long is an acceptable amount of time to wait for opt-out messages before moving forward with data collection? Also, what happens if users do not see the message right away, and will individual users that have issued opt-out requests trust researchers that remain in the group chat to continue monitoring only the other chat messages? Also, announcing researcher presence could have a chilling effect on the natural discussion of the group chats~\cite{matias2020automated}, potentially affecting both the quality of accurate data collection and regular WhatsApp users' legitimate experiences. 
Moreover, anonymity has notable limitations for participant privacy. Data can be ``de-anonymized'' by malicious or fallible agents with access to the cross-reference or from other attacks such as data reconstruction (e.g.,\ US Census data reconstruction~\cite{bureau2021disclosure}). 
While some privacy frameworks such as differential privacy go beyond anonymity by adding mathematically constructed noise to protect against such attacks, these methods have not been used in existing WhatsApp research.
Overall, WhatsApp users' oversight in data collection is largely excluded, justified by the data being publicly accessible. Our work builds on prior approaches to WhatsApp data collection to include users more actively in this process. 

\subsection{Relevant Privacy Frameworks}
The most popular privacy frameworks relevant to this work include Privacy by Design~(PbD)~\cite{cavoukian2009privacy} and Data Minimization~(DM)~\cite{biega2020operationalizing}, but these are broad guiding principles for product designers with heterogeneous interpretations~\cite{biega2020operationalizing, van2014designing} as opposed to the methodological design considerations for research-driven data collection that we put forth in this work. 
The most relevant contemporary privacy framework to this work is Contextual Integrity (CI). From a CI perspective~\cite{nissenbaum2019contextual, nissenbaum2004privacy}, common methods of WhatsApp data collection compromise participant privacy by violating the contextual norms surrounding online group chats.
In fact, Internet researchers more broadly have found that collecting public user data is fraught with ethical/methodological pitfalls because contextual norms are violated~\cite{zimmer_but_2010, cooper_ethics_2020, hagen_all_2020, mancosu_what_2020, fiesler_participant_2018, vitak_ethics_2017, verheijen_collecting_2016, tranberg_dataethics_2018}. 
For example, 
Zimmer found ``considerable conceptual gaps in the understanding of the privacy implications of research in
social networking spaces''~\cite{zimmer_but_2010} with respect to publicly released Facebook user data. Researchers have also found a lack of consensus among Internal Review Board (IRB) staff regarding standards for computational social science research, including ``the challenge of obtaining informed consent in large-scale data projects''~\cite{vitak_ethics_2017}. To the best of our knowledge, our work is unique in seeking user input regarding WhatsApp data collection methodologies.

\section{Study Part~1: Design Considerations and Feasibility}
\begin{figure*}
    \centering
    \begin{subfigure}[t]{0.19\textwidth}
        \includegraphics[width=\textwidth]{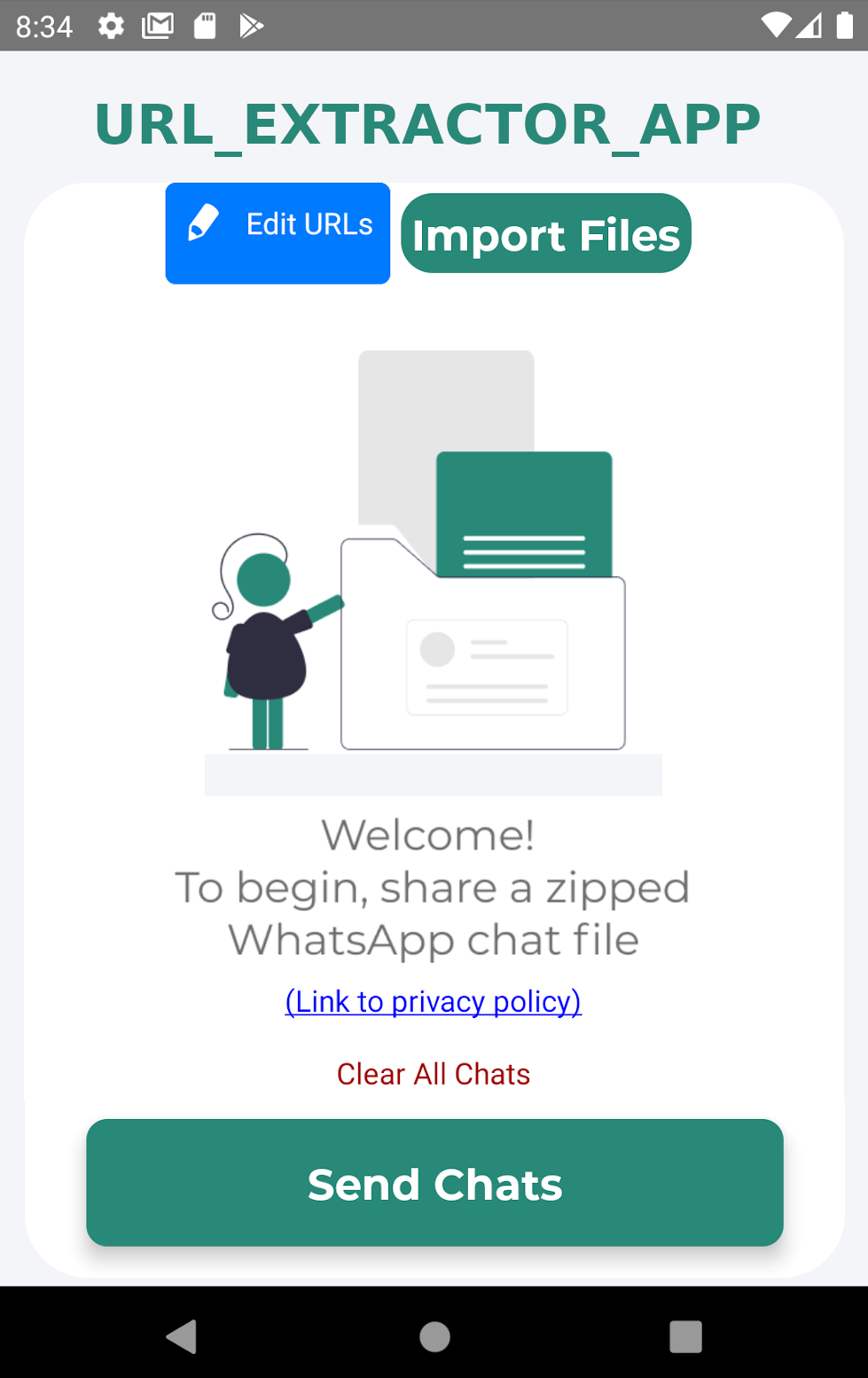}
        \caption{}
        \label{fig:apphome}
    \end{subfigure}
    \hfill
    \begin{subfigure}[t]{0.19\textwidth}
        \includegraphics[width=\textwidth]{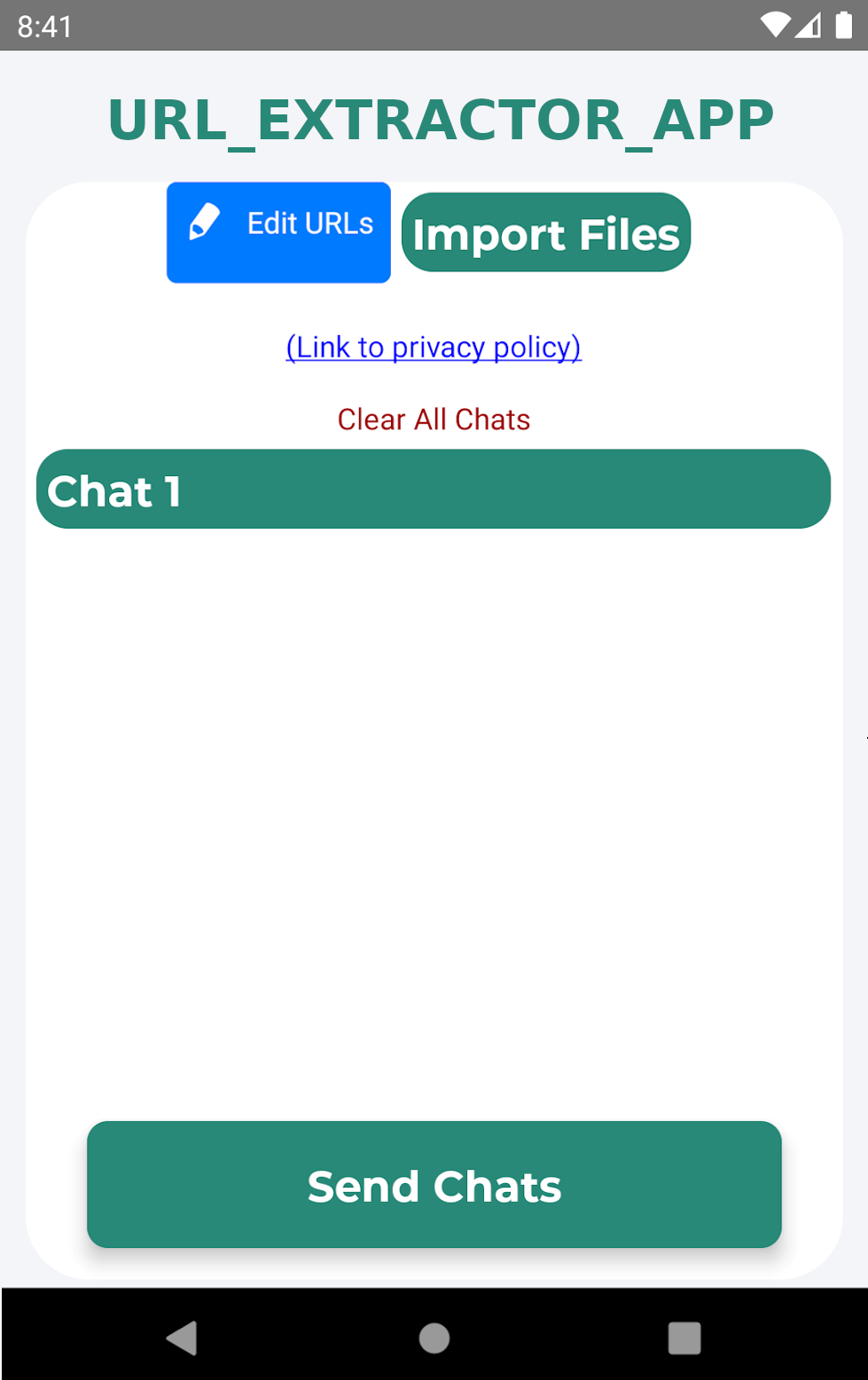}
        \caption{}
        \label{fig:appchatlist}
    \end{subfigure}
    \hfill
    \begin{subfigure}[t]{0.19\textwidth}
        \includegraphics[width=\textwidth]{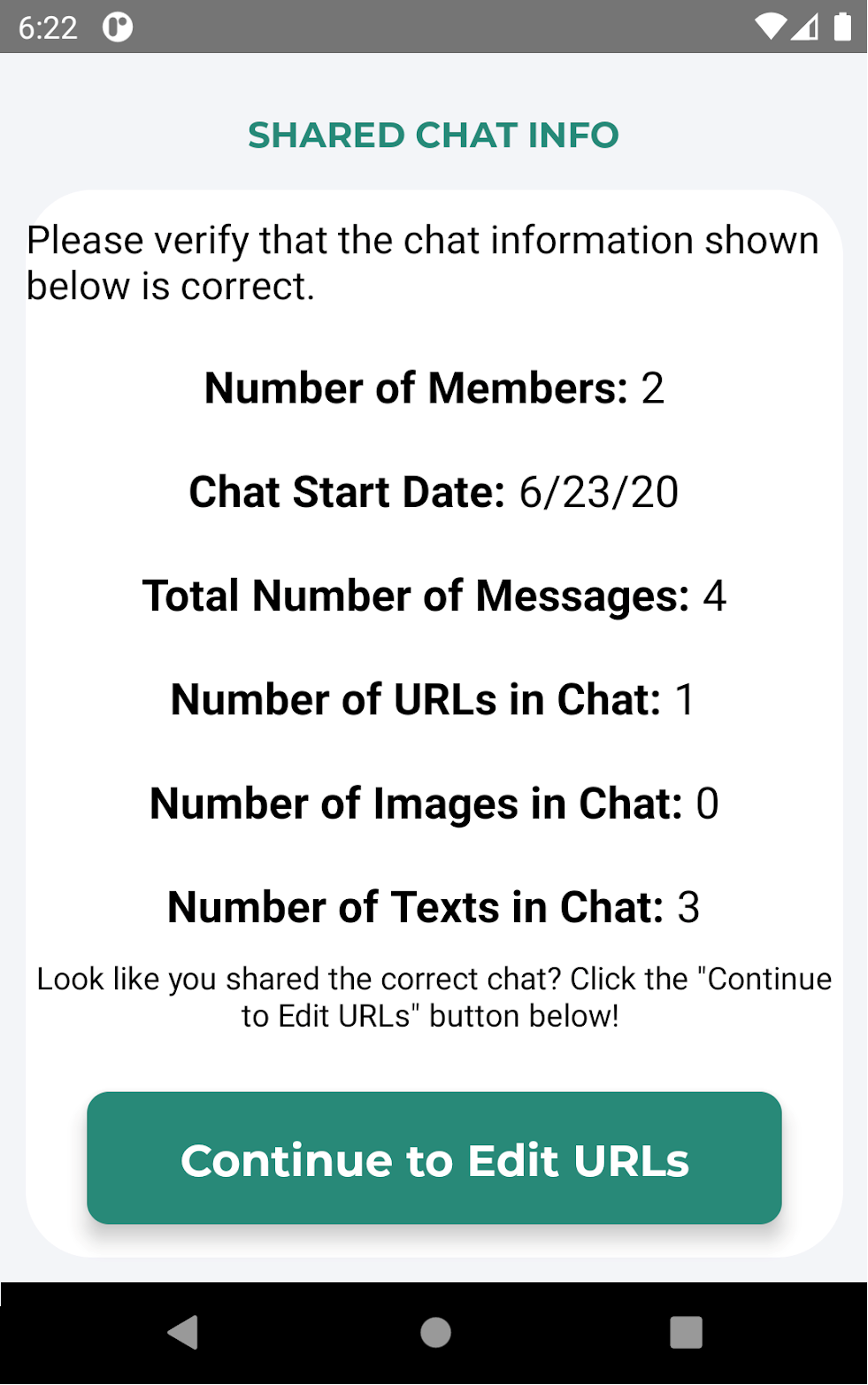}
        \caption{}
        \label{fig:appcharinfo}
    \end{subfigure}
    \hfill
    \begin{subfigure}[t]{0.19\textwidth}
        \includegraphics[width=\textwidth]{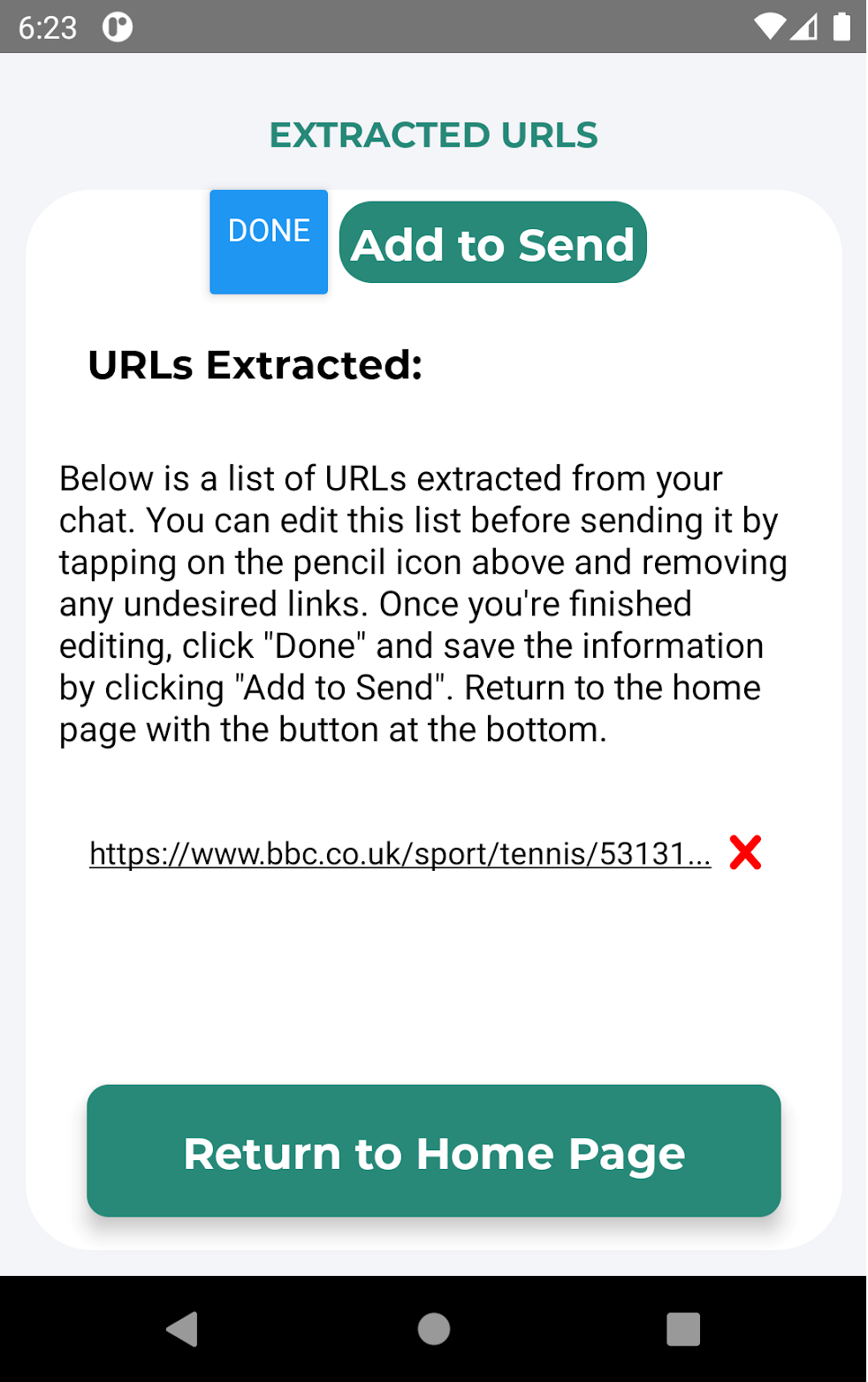}
        \caption{}
        \label{fig:appediturls}
    \end{subfigure}
    \hfill
    \begin{subfigure}[t]{0.19\textwidth}
        \includegraphics[width=\textwidth]{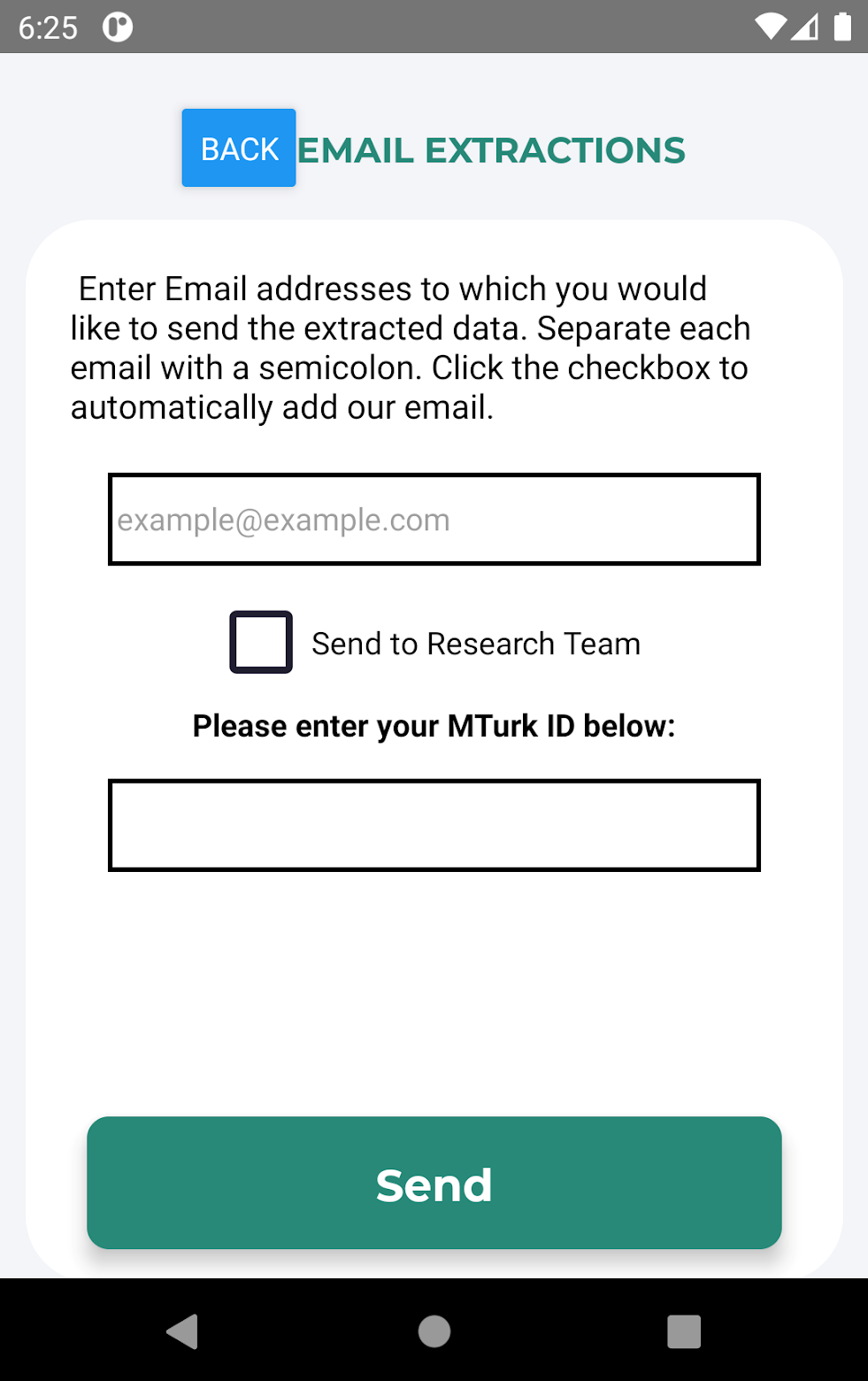}
        \caption{}
        \label{fig:appshare}
    \end{subfigure}
    \caption{Screenshots of URL-EXTRACTOR-APP's navigation screens. A user could either add an exported WhatsApp chat text file or share the file directly to the app (a). Once the exported chats were shared to or imported into URL-EXTRACTOR-APP, the app displayed a list of all shared chats. It is shown as `Chat 1' in this example (b). The example metadata for a chat could be verified (c). The list of URLs for the selected chat was displayed as a list with an `X' next to it (d). Clicking this `X' deleted the URL from the chat's metadata. When a user clicked `Add to send', the user could then enter the email addresses for which to send the extracted chat along with the ability to automatically send it to the research team (e).}
    \label{fig:screenshots}
\end{figure*}

To answer RQ1, we first created four UCDS principles for collecting WhatsApp chat data in a way that gives users more control over what data they share and encourages researchers to minimize the data they collect and process off a user's device.
To answer RQ2 and demonstrate the feasibility of deploying UCDS principles for research purposes, we then implemented and evaluated one possible instance of UCDS principles in the form of a mobile application with a small-scale study: URL-EXTRACTOR-APP. The app design and usage by a user is described in Figure~\ref{fig:screenshots}.

\subsection{User-Centered Data Sharing Principles}
\label{sec:privacymech}

Here we introduce the UCDS principles in more detail and use the development of URL-EXTRACTOR-APP as a recurring example implementation of each principle.
The principles build upon common practices mentioned in Section~\ref{commonpractices} to also include: \textit{Constrained Data Collection}, \textit{Local Extraction and Processing}, \textit{User Involvement}, and \textit{Transparency}.
The full UCDS principles are detailed below. In practical implementation, the principles often work hand in hand. 

\textbf{Constrained Data Collection}
To maximize understanding of the content contained within private chats, it is tempting to collect all shared content: images, texts, videos, documents, audio, and metadata. However, this data is likely to contain PII typically exceeds the amount of data required for research studies. Following \textit{Constrained Data Collection} means only targeting data that can be collected using privacy preserving methods and narrowing the data collection scope to strictly necessary data for each study.
For example, URL-EXTRACTOR-APP did not extract the content of any message type unless it was a URL. 
The app replaced usernames with coded identifiers (e.g.,\ `User1') so that we could track how many people were in a chat and which users were sending messages without recording identities. 
URL-EXTRACTOR-APP extracted only the following data:
\begin{itemize}
    \item The start and end dates of the chat
    \item The number of users in the chat
    \item The number of messages from each user in a chat overall
    \item The number of messages from each user per day
    \item The number of URLS and text messages from each user
    \item The message type (URL or text) of each message
    \item The URLs in each chat
    \item The anonymized sender of each message (e.g.,\ John Doe and Jane Doe become User0 and User1, respectively)
    \item The date each message was sent
\end{itemize}
The data we collect is similar to the metadata that Meta collects from all WhatsApp chats~\cite{dhapola_whatsapp_2017}. 

We designed the app to focus on URLs for three reasons. First, prior work has shown that misinformation (a primary area of focus for WhatsApp researchers) often spreads in the form of URLs~\cite{narayanan_news_2019, machado_study_2019}. Second, shared URLs may operate as a proxy for chat content, providing a glimpse into a chat's discussion without reading the private messages explicitly. Third, since images, texts, audio, and video being sent can contain PII that is hard to scrub, we felt that participants would be more comfortable with letting us collect URLs from their chats. Upon collection, we stripped the URLs down to their domain name and top-level domain (TLD) due to sensitive information that could be gathered from the full-length URL, such as their online usernames or events they had attended. For example, shared links to an online Zoom meeting or YouTube video became simply ``zoom.us'' and ``youtube.com''. We also kept country code top-level domains (ccTLD) and followed redirections from URL shorteners. By reducing links to their domain-level, we minimized the risk of violating the participants' privacy while still being able to characterize the types of links shared. Even the constrained data collected by URL-EXTRACTOR-APP is more than necessary for some research questions; it exemplifies types of minimally-invasive yet useful chat data. 

\textbf{Local Extraction and Processing}
In UCDS, filtering the available data down to the constrained data and performing any computations should occur locally on the users' devices insofar as it is possible.
Following \textit{Local Extraction and Processing} gives participants full control over the extraction process and the extracted data before any information is transferred off the device. Thus, the risk of transferring potentially private information is minimized. 
In our URL-EXTRACTOR-APP, to implement this UCDS principle, we deployed our app to study participants so they could launch the data extraction from their WhatsApp chats locally.
In the case of URL-EXTRACTOR-APP, data collection via \textit{Local Extraction} removed the need for our research team to join any WhatsApp chats, which is the primary means of public WhatsApp data collection and introduces additional privacy concerns. Instead, the users could export their WhatsApp chats and share them directly to URL-EXTRACTOR-APP.
URL-EXTRACTOR-APP analyzed the file, packaged the above data, and anonymized it fully locally. 

\textbf{User Involvement} 
The principle of \textit{User Involvement} seeks to balance the data collection relationship between researchers and participants. Involving users is enabled by conducting the extraction and data processing locally giving users oversight and final say when it comes to what data is shared. In our implementation example, URL-EXTRACTOR-APP allowed users to manage the data extraction and choose whether to continue with sharing their chat data, on a chat-by-chat basis. Specifically, users were shown the list of URLs and metadata that the URL-EXTRACTOR-APP extracted from each respective WhatsApp chat and were able to remove any URLs from the list that they did not want to share (Figure~\ref{fig:appediturls}). As such, these URLs were not added to our dataset. However, we recorded whether the list of a chat's URLs had been edited or not because we still wanted to know whether the extracted list of URLs was all or a subset of the shared URLs. We also wanted to know whether participants would edit the extracted data or share it with us as is in our feasibility study.

\textbf{Transparency} 
Following the principle of \textit{Transparency} equips users with the information they need in order to make informed decisions about sharing their data. \textit{Transparency} covers both the study details and `observation windows' into the data collection procedures. For our implementation of URL-EXTRACTOR-APP, we provided detailed documentation at ANONYMIZED-URL to users on how the application was to be used, how it operated, and the motivations of our study through a privacy policy sheet with frequently asked questions. 
We also ensured that users could view the extracted data in its complete form by offering the option for them to email the extracted chat file to anyone (including themselves) as well as the research team. Participants were also able to uncheck a box that sends the email to the research team if they wanted to preview the file prior to sending it to the team.

\subsection{App Implementation and Testing}
We coded URL-EXTRACTOR-APP in JavaScript using the React Native framework and published it to the Apple App store. To evaluate whether our specific implementation of the UCDS principles (URL-EXTRACTOR-APP) was a feasible means of collecting WhatsApp chat data, we conducted a small-scale deployment study. Our user study was approved by the IRBs of our two institutions who also checked that URL-EXTRACTOR-APP did not violate WhatsApp's terms of service, chat member's rights, or wiretapping laws in our state. We recruited participants using social media at both of our institutions in the US and a recruitment survey center at one of our institutions between May and June 2021. Each participant completed a demographic survey (see Appendix~\ref{bsi}) and  reported their WhatsApp usage.

For each participant, we conducted a Zoom session that lasted 15--45 minutes where we guided participants through using the application to share chat data with the research team. 
We left the number of chats to share up to the participant. We asked that they shared chats they participated in the most or ones with high amounts of shared links. After the participants shared data with us, we thanked them and concluded the session. Each participant was compensated with a \$15 gift card. 

\textbf{Data Analysis}
We analyzed the dataset of private WhatsApp chat data for chat member dynamics such as how many members were in each chat and who sent each link. Next, we looked at the emergence of URLs in the dataset by calculating the frequency and occurrences of links being shared relative to text only messages. Third, we analyzed what type of information was being shared by analyzing the links' domains. We also analyzed the TLDs and ccTLDs of each link. In the dataset, the first user sending a message in a chat is User0, incrementing the counter for each subsequent unique user---User1, User2, etc. For each chat per participant, we label the first chat as A, the second as B, and so on.

The participants had an uneven representation in the dataset because they shared different numbers of chats which contain different amounts of messages. To avoid skewed data in our analysis, we report medians. To calculate overall medians, we first calculated medians across all chats for each participant. Then, we calculated medians of the participants' medians, to not skew data analysis towards over-represented participants. For example, to calculate the percentage of messages that contained URLs, we first calculated the median percentage of messages containing URLs for all a participant's chats. Then, we take the median percentage across all participants.

\begin{table}
\caption{A summary of the participants and dataset gathered from the UCDS feasibility study, including the number of chats collected by our app, how long they have used WhatsApp, their number of WhatsApp contacts, the total number of URLs in their chats, and the median durations of their chats. }
\centering
\resizebox{0.92\columnwidth}{!}{%
\begin{tabular}{|l|r|r|r|r|r|}
\hline
\textbf{Participant} &
  \multicolumn{1}{c|}{\textbf{\# Chats}} &
  \multicolumn{1}{c|}{\textbf{\begin{tabular}[c]{@{}c@{}}\# Years Used\\ WhatsApp\end{tabular}}} &
  \multicolumn{1}{c|}{\textbf{\begin{tabular}[c]{@{}c@{}}\# WhatsApp\\ Contacts\end{tabular}}} &
  \multicolumn{1}{c|}{\textbf{\begin{tabular}[c]{@{}c@{}}\# URLs in\\ Chats\end{tabular}}} &
  \multicolumn{1}{c|}{\textbf{\begin{tabular}[c]{@{}c@{}}Median Chat\\ Len. (months)\end{tabular}}} \\ \hline
P1                          & 3   & 7  & 105   & 194  & 20.6 \\
P2                          & 5   & 10 & 150   & 152  & 67.9 \\
P3                          & 5   & 5  & 400   & 140  & 26.4 \\
P4                          & 5   & 7  & 25    & 20   & 4.9  \\
P5                          & 3   & 5  & 82    & 71   & 13.7 \\
P6                          & 4   & 7  & 70    & 18   & 19.6 \\
P7                          & 1   & 7  & 70    & 30   & 3.7  \\
P8                          & 3   & 10 & 500   & 110  & 4.6  \\
P9                          & 4   & 3  & 24    & 23   & 1.1  \\
P10                         & 3   & 3  & 118   & 336  & 15.9 \\ \hline
\multicolumn{1}{|r|}{Median} & 3.5 & 7  & 111.5 & 90.5 & 14.8 \\ \hline
\end{tabular}%
}

\label{tab:Pdata}
\end{table}

\textbf{Participants}
10 participants in total used URL-EXTRACTOR-APP to share chat data with us. Our participants were younger, and many had higher education degrees reporting varied incomes. Five participants were between ages 18-24 and five were 25-35 years old. Four had a bachelor's degree, five had a master’s degree and only one reported having only some college. Six were female identifying and the rest were male identifying. Five reported making \$50K or less, three reported making >\$50K and two declined to answer. The number of chats for which participants used our app to automatically extract data and shared with us ranged from one to five per participant, with a median of 3.5. In total, participants shared data from 36 chats.  We denote participants in Study Part~1 using the notation P1--P10. The exploratory dataset arising from the feasibility deployment is summarized in Table~\ref{tab:Pdata}. 
It consists of a wide range of chat lifetimes, from a six-day chat (P8) to an almost six-year chat (P2; created in June of 2015). The median chat length in our dataset was 1.23 years, and the median number of URLs shared in a chat was 15.

\subsection{Study Part~1 Findings: UCDS Is Feasible}
Study Part~1 demonstrates the ability to garner insights following the UCDS principles. We use Garimella and Tyson as a recurring point of reference~\cite{garimella_whatsapp_2018}, who conducted an early feasibility study of using WhatsApp data for research collected from \textit{joining and scraping} public chats. Note, our user study resulted in a small-sized exploratory dataset and serves as a proof-of-concept evaluation for methodologies following the User-Centered Data Sharing principles; these findings may be sample dependent. We had three main insights from the URL-EXTRACTOR-APP suggesting that research using UCDS principles is not prohibitive for deriving research insights.

\textbf{Participant Chat Dynamics.} Over 80\% of the chats in our dataset had only two users, but there were several larger private chats with as many as 25 users. The number of users for every chat in our dataset is presented in Appendix~\ref{sec:extra:table}. Related work has found public chats to be much larger, with a median of 127 members~\cite{garimella_whatsapp_2018}.
Regardless of the chat size, a sole user was responsible for a majority of the links shared in the private chats we examined over the course of the chat's lifetime. 
Across the 36 chats, the single user that sent the most links was responsible for 64\% of the links of a chat (including chats with more than two members).
 Prior work where researchers joined and scraped public chats found similar but less extreme skews towards a small portion of disproportionately active users~\cite{garimella_whatsapp_2018}.

 \begin{table}
\caption{The total number of messages in our dataset, including the number of shared links and text-only messages. }
\centering
\resizebox{0.4\columnwidth}{!}{%
\begin{tabular}{|l|r|}
\hline
\textbf{\# URLs} & 1094 \\ \hline
\textbf{\# Texts} & 112,523 \\ \hline
\textbf{Total Messages} & 113,617 \\ \hline
\end{tabular}
}
\label{tab:linktext}
\end{table}
 
 \textbf{Trends of Relatively Low URL Exposure.} The median number of times a domain was shared throughout a chat's lifetime in our dataset was one, resulting in a distribution containing a short peak of higher-shared domains followed by a long tail of  domains shared infrequently. Overall, URLs were uncommon in our dataset of private chats and dwarfed by text messages (Table~\ref{tab:linktext}). By median of medians, only 1.09\% of all messages contained links, a much smaller portion than what previous literature has shown in public group chats~(39\%~\cite{garimella_whatsapp_2018}).
     
\textbf{Varied Types of Shared Links.} As listed in Table~\ref{tab:tld}, the domains that were the most shared across the 36 chats were \textit{YouTube} (4.4\%), \textit{Google} (4.2\%), and \textit{Twitter} (0.8\%). These three were the only domains that appeared in at least half of the chats for at least half of the participants. Notably, Garimella and Tyson found \textit{YouTube} to be the most popular domain in their dataset of public WhatsApp chats as well~\cite{garimella_whatsapp_2018}. 44\% percent of chats contained links with international country codes (e.g.,\ `.mx' (Mexico) and `.fr' (France)). Note, we excluded ccTLDs being used for aesthetic or brand appeal (e.g.,\ `twitch.tv'). Table~\ref{tab:tld} shows all the ccTLDs present in the dataset. Prior work also found that 85\% of public chats had members representing over 10 countries~\cite{garimella_whatsapp_2018}.

\subsection{Feasibility and Limitations}
Our findings provide examples of how using UCDS methods for WhatsApp chat data collection can still assist in answering a variety of research questions. The many WhatsApp studies that operate by automatically joining and scraping chats, which violate UCDS principles, use similar insights to answer their research questions (albeit at larger scales). Study Part~1 suggests that maintaining heightened participant privacy with UCDS methods is feasible for research purposes. 
There are several limitations for using the development and deployment of URL-EXTRACTOR-APP as a feasibility study for the UCDS principles. For example, guiding participants through the app usage in Zoom meetings was resource intensive, limiting the dataset to a smaller scale. 
Also, the full reach of UCDS principles is difficult to measure from a feasibility study in a single context. 
Future studies in other contexts may highlight unforeseen parameters and require adjustments to the design of the data sharing mechanisms. Additionally, the UCDS principles can be expanded upon to fit study needs and learned user preferences as further elaborated in Section \ref{Discussion}. 

Finally, recall that \textit{User Involvement} allowed for users to edit their chat data before we could analyze it. While we consider this a privacy enhancing feature---that may increase user participation and thus data collection volume in the long run---as opposed to a true limitation, it is worth noting that providing this ability has the potential to skew takeaways. In Study Part~1, only 3/10 participants edited their chats totaling 8/36 edited chats.

\begin{table}[t]
\centering
\caption{The domain properties in our dataset across all 36 chats, including the most popular domain names, most popular TLDs, and all ccTLDs present. }
\resizebox{0.75\columnwidth}{!}{%
\begin{tabular}{|c|c|c|}
\hline
\textbf{Top Domains} & \textbf{ccTLDs} & \textbf{Top TLDs} \\ \hline
youtube (4.4\%) & \multirow{3}{*}{\begin{tabular}[c]{@{}c@{}}{[}.mx, .uk, .us, .fr, .es, \\ .eu, .ru, .nl, .py, \\ .pk, .hu, .ca, .in{]}\end{tabular}} & .com (75.7\%) \\
google (4.2\%) & & .org (4.5\%)\\
twitter (0.8\%) & & .edu (1.2\%) \\ \hline
\end{tabular}
}
\label{tab:tld}
\end{table}

\section{Study Part~2: User Perceptions of WhatsApp Data Collection Methods}
The findings of Study Part~1 demonstrated the feasibility of following the heightened privacy-persevering UCDS principles for collecting WhatsApp chat data and showed that insights about the private chats’ contents can be gleaned such as what type of information was being shared, who shared it, and how often it was shared. However, this study did not examine user perceptions about the different methods for collecting WhatsApp chat data or UCDS principles (RQ3). We deployed a survey in Study Part~2 to answer RQ3 and better understand user expectations of privacy and contextual norms when researchers are involved in WhatsApp chats.

\subsection{Survey Design}
\begin{figure}[t]
    \begin{minipage}[t]{0.45\textwidth}
        \centering
        \begin{framed}
        \caption*{\normalsize Scenario 1: Standard Method}
        \caption*{\scriptsize Imagine the following scenario: \textbf{The invite link to one of your WhatsApp chats is publicly available online, and a researcher uses it to join the chat.}}
        \vspace{-1pt}
       \includegraphics[width=0.8\columnwidth]{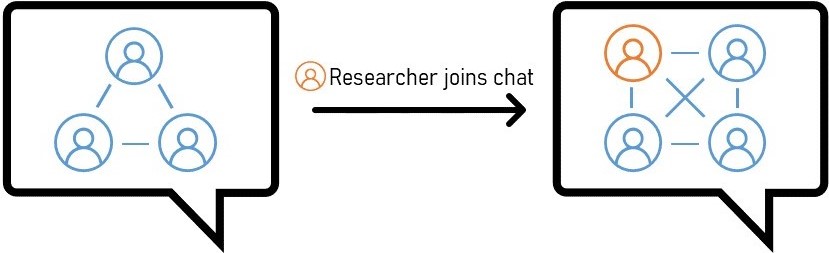}
        \end{framed}
        \caption{Illustration provided to participants explaining the standard methods for collecting WhatsApp chat data where researchers join WhatsApp chats via invite links found online.}
        \label{fig:join}
    \end{minipage}
    \begin{minipage}[t]{0.45\textwidth}
        \centering
        \vspace{12pt}
        \begin{framed}
        \caption*{\normalsize Scenario 2: UCDS Principles}
        \caption*{\scriptsize Imagine the following scenario:\textbf{ One of your WhatsApp contacts is involved in a research study that is collecting chat data, including data from a chat that you are also in. In this scenario, the researchers do not join your chat but are in contact with one of the chat members.}}
        \vspace{-1pt}
         \includegraphics[width=0.8\columnwidth]{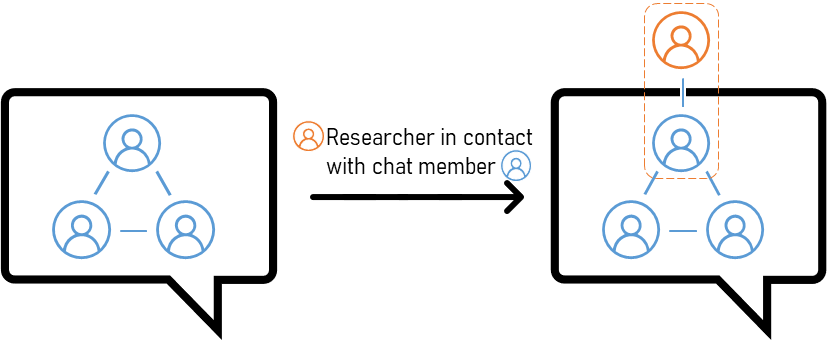}
        \end{framed}
        \caption{Illustration provided to participants explaining the scenario representing UCDS. Specifically, researchers do not join the chats but externally interface with a chat member (e.g.,\ via an application like URL-EXTRACTOR-APP).}  \label{fig:notjoin}
    \end{minipage}
\end{figure}

To gather user perceptions of WhatsApp data collection methods including UCDS, we designed a survey grouping questions into the following topics:\footnote{The full survey instrument can be found in Appendix~\ref{user_perceptions_survey}.}
\begin{itemize}
    \item \textbf{WhatsApp Usage:} After providing study information and obtaining consent, we asked each participant about their general WhatsApp usage and contacts.
    \item \textbf{Data Sharing Comfort by Data Type:} We then asked participants to indicate the comfort level they felt towards sharing different categories of chat data with researchers on a 5-point Likert scale (Extremely Comfortable to Extremely Uncomfortable). We included a range of data categories from minimally invasive data categories (e.g.,\ \textit{the number of users in the chat}) to PII (e.g.,\ \textit{the names of all chat members}). The data categories were presented in a random order while keeping similar categories next to each other to reduce cognitive burden on the participants (e.g.,\ \textit{whether a message contains a URL} and \textit{the URL itself} would always be adjacent). When presenting results, we categorize the different data types into Metadata, Message Contents, and PII.
    \item \textbf{Data Collection Scenarios:} We then asked participants to answer questions based on two hypothetical data collection scenarios: one representing current techniques and one representing the UCDS principles from Study Part~1. In the first scenario, a researcher joins a WhatsApp chat via a publicly accessible invite link to collect the chat's data (Figure~\ref{fig:join}). Since the invite link is publicly available online, we consider the hypothetical chat in Scenario 1 a \textit{public} chat~\cite{feng_investigating_2022}.
    In the second scenario, a researcher is in contact with a chat member but does not join the WhatsApp chat (Figure~\ref{fig:notjoin}). Chats in Scenario 2 can be either private or public chats.
    URL-EXTRACTOR-APP (from Study Part~1) is one possible instance of Scenario 2. Separating the researchers from the chat allows for additional questions about how data filtering and anonymization can occur in ways that are not possible in contemporary methods of joining and scraping chats. Participants answered 5-point Likert-scale questions related to both scenarios.
    The two scenarios were presented in a random order, and each concluded with an open-ended question to provide any additional thoughts about the scenario for scraping WhatsApp chat data.
    \item \textbf{Miscellaneous:}
    Next, we asked participants a few miscellaneous questions related to WhatsApp data collection such as their willingness to share data for research purposes and trust levels towards their contacts and researchers. 
    \item \textbf{Demographics:} Lastly, we gathered demographic information including age, education, and gender identity and asked participants to share any final open-ended thoughts about WhatsApp data collection in general. 
\end{itemize}

\begin{table*}[t]
\centering
\caption{Demographics of Study Part~2: age, education, gender, and annual income. }
\label{tab:demogs2}
\resizebox{0.75\textwidth}{!}{%
\begin{tabular}{|lrr|lrr|lrr|lrr|}
\hline
\textbf{Age} & \textbf{\#} & \textbf{\%} & \textbf{Education}              & \textbf{\#} & \textbf{\%} & \textbf{Gender}      & \textbf{\#} & \textbf{\%} & \textbf{Income}         & \textbf{\#} & \textbf{\%} \\ \hline
18-24        & 39          & 12          & High School                     & 22          & 7           & Female Identifying   & 168         & 50          & \textless{}\$25,000     & 67          & 20          \\
25-34        & 144         & 43          & Some College      & 56          & 17          & Male Identifying     & 161         & 48          & \$25,000-\$50,000         & 86          & 26          \\
35-44        & 90          & 27          & Associate Degree                & 30          & 9           & Non-Binary           & 2           & 1           & \$50,000-\$100,000        & 112         & 34          \\
45-54        & 34          & 10          & Bachelor's Degree               & 154         & 46          & Pref. not to answer & 3           & 1           & \$100,000-\$200,000       & 53          & 16          \\
55+          & 27          & 8           & Grad.\ or Prof.\ Degree & 70          & 21          &                      &             &             & \textgreater{}\$200,000 & 9           & 3           \\
             &             &             & Other                           & 2           & 1           &                      &             &             & Pref. not to answer    & 7           & 2           \\ \hline
\end{tabular}%
}
\end{table*}

\subsection{Survey Deployment and Analysis}
We built the survey using Qualtrics~\cite{noauthor_qualtrics_nodate} and piloted the survey with volunteers familiar with HCI research to refine it for clarity and flow. We received IRB approval and deployed the survey to 350 participants recruited via Prolific~\cite{noauthor_prolific_nodate}. We filtered for participants currently located in the US that regularly use WhatsApp. 
We paid participants based on ANONYMIZED-STATE minimum wage for 20 minutes of work (based on piloting response times). We excluded a total of 16 responses for either failing one of the survey's two attention check questions or not qualifying for the study (e.g.,\ a non-WhatsApp user), resulting in 334 total responses. We denote participants in Study Part~2 using the notation S1--S334. When reporting Likert results, we bin together \textit{strongly} and \textit{somewhat} responses. That is, if a question had a response breakdown of 44\% Strongly Disagree, 21\% Somewhat Disagree, 15\% Neutral, 15\% Somewhat Agree, and 5\% Strongly Agree, we would report this as 65\% of participants disagreed and 20\% agreed with the prompt, the rest being neutral. 

\textbf{Survey Participants}
The survey participants' age distribution approximately matched the US WhatsApp population~\cite{ceci_us_2022}, and most participants (67\%) had completed at least a bachelor's degree reporting a variety of incomes. The full participant demographics for Study Part~2 are shown in Table~\ref{tab:demogs2}. The median number of years participants used WhatsApp was 5 years with a range of less than a year to 13 years maximum.\footnote{WhatsApp was first available in 2009, 13 years prior to the time of survey deployment.} The participants had a median of 30 WhatsApp contacts. The minimum number of contacts reported was one, and five participants checked a box indicating they had more than 500 WhatsApp contacts.

\subsection{Study Part~2 Findings: UCDS Is Preferred But Needs Improvement}
From the survey of 334 WhatsApp users, we found that the UCDS principles benefit users by putting them in control of the data collection process and identified aspects that future iterations of best practices could address. Specifically, the survey findings are summarized into three main themes: we uncovered concerns with current approaches where researchers join and scrape public WhatsApp chats; we found that following the UCDS principles to collect WhatsApp chat data is well-suited for implementing users' privacy preferences;
and complexities related to group consent remain unsolved by the UCDS principles in their current form.
We elaborate on these themes in the following sections.

\begin{figure*}[t]
    \centering
    \includegraphics[width=.85\textwidth]{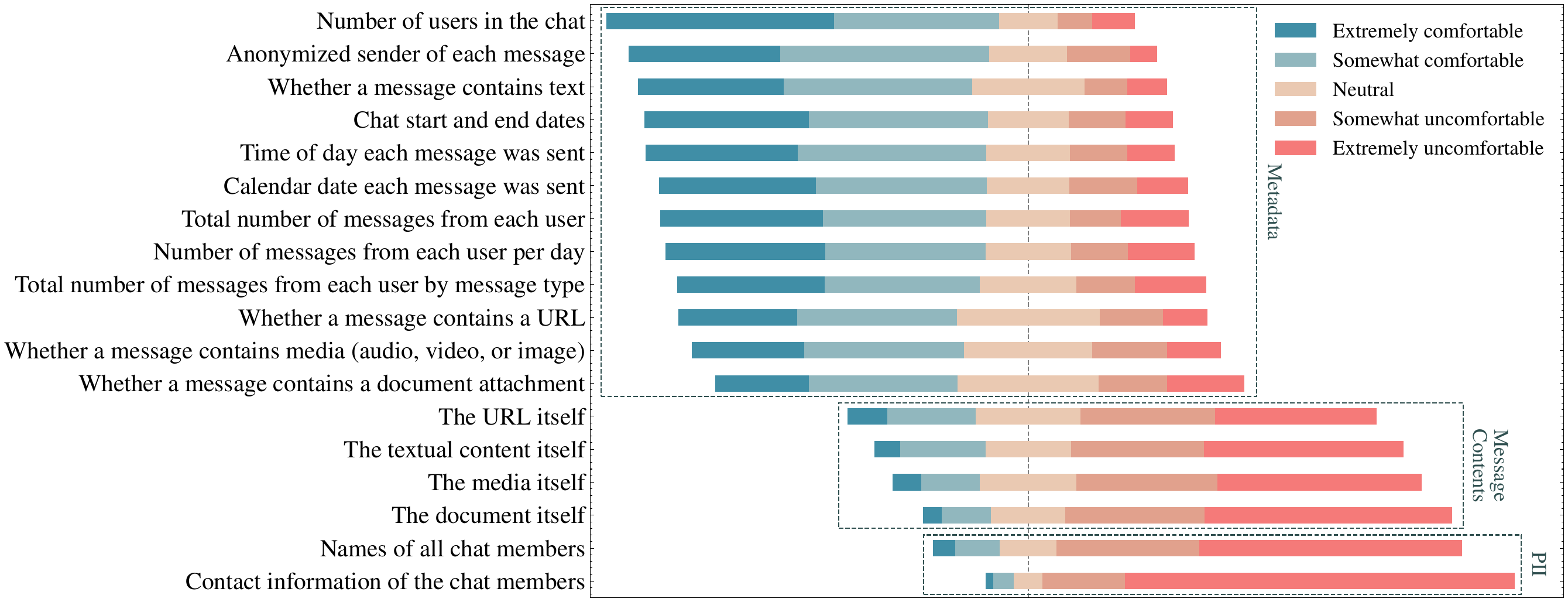}
    \caption{Participant comfort level sharing different types of WhatsApp chat data, grouped by Metadata, Message Contents, and PII.}
    \label{fig:datashare_comfort}
\end{figure*}

\subsubsection{Concerns With Scraping Public WhatsApp Chats}
The survey respondents indicated that most of their WhatsApp chats were private chats with close contacts where no invite link was publicly available, and that they were concerned about researchers collecting their chat data without explicit consent even if the chat data were publicly accessible. 

\textbf{Most Whatsapp Chats Are Not Accessible by Online Public Invite Links.}
Survey respondents indicated that the majority of their WhatsApp conversations were private chats among family and friends; suggesting that research targeting only publicly accessible WhatsApp chats~\cite{garimella_whatsapp_2018, javed_first_2020, machado_study_2019, maros_analyzing_2020, narayanan_news_2019, resende_analyzing_2019, resende_misinformation_2019, saha_short_2021, garimella_images_2020, melo_whatsapp_2019, moreno_whatsapp_2017, reis_dataset_2020, shah_tools_2019} misses a lot of the WhatsApp ecosystem. When asked to characterize their WhatsApp contacts, most participants (92.8\%) said they had contacts that were friends, followed by 78.7\% that selected family (participants could select multiple responses). One third (32.6\%) had contacts that were work colleagues, and only 15.3\% said they were contacts with people they only know from WhatsApp group chats. Similarly, the most common ways that participants joined WhatsApp chats were from their friends and family directly adding them to chats (93.7\%) or by creating their own chats and adding their contacts (63.7\%). Only 11.7\% selected that they joined WhatsApp chats by public online invite links. 

\textbf{Users Do Not Want Data to Be Collected From `Public’ Whatsapp Chats Without Their Consent.}
When asked about researchers joining a public chat to collect WhatsApp chat data (Scenario 1), almost all participants (92.5\%) agreed that even if an invite link was publicly accessible online, that researchers should inform the members of their presence and obtain consent from all chat members before collecting data. Only 3\% disagreed and 4.5\% were neutral. Some participants provided more detail about their reasoning, for instance:
    ``\textit{I do understand that `publicly available' generally means accessible to anyone, and that makes sense, but I still feel it would be best practice to inform everyone involved of the intention to collect data, even if that could compromise or alter behavior}''~(S27), and 
    ``\textit{Groups of friends talk in public and its still considered rude if a stranger interjected into a conversation. Then took a picture of you}''~(S248).

These results suggest that users would like researchers to announce presence and intent even if the chat is technically publicly accessible.
Respondent S49 gave context to the dissenting opinions which aim to preserve the original nature of researched chats:
    ``\textit{If researchers inform their intention, the dynamic of the group chat might change and people might act differently and unnaturally because they feel uncomfortable being studied or observed}.'' Moreover, 83.5\% of participants agreed that their willingness to share their WhatsApp data with researchers depended on the goals of the research (9\% were neutral and 7.5\% disagreed). For users to exercise their right to not participate in research studies, they may want to know more about the research. S83 explained:
    ``\textit{I wouldn't want them creeping in on my work conversations. [...] Like work documents I don't think I would be to comfortable with sharing, unless I knew what the research was for}.'' 
The high degree of circumstance influencing comfort has also been shown for Twitter~\cite{fiesler_participant_2018} and Facebook~\cite{gilbert2021measuring} users.

\subsubsection{UCDS Principles are Well-Suited for Implementing Users’ Privacy Preferences}
Our survey respondents reported that WhatsApp data should be filtered down to just necessary metadata, anonymized, and given participant oversight before researchers could even access this data. These preferences are made possible by Scenario 2 and lend themselves to a method following the UCDS principles of Study Part~1 where data extraction happens locally in the hands of WhatsApp users.

\textbf{Data Collected Should Be Limited to Metadata If Possible.} 
When we asked participants about their comfort level sharing the various types of data associated with any WhatsApp chat, they were much more comfortable sharing metadata over the actual chat contents. As shown in Figure~\ref{fig:datashare_comfort}, there is a steep decrease in comfort when participants were asked about sharing a message's content itself versus the type of content contained in a message (i.e., text, URL, media, or document) or other high-level chat metadata like the date the chat started.
Further, of actual message contents we asked about collecting, participants were least uncomfortable sharing messaged URLs.
Participants were also most uncomfortable with sharing names or contact information with researchers.
In sum, participants preferred that only (and anonymous) metadata be extracted as opposed to the full chat data. In contrast, common methods of WhatsApp data collection scrape full chat data, sometimes including member identities~\cite{garimella_images_2020, melo_whatsapp_2019, moreno_whatsapp_2017, shah_tools_2019}. 
In an open-response survey question, S15 concisely summarized this approach: ``\textit{All images and audio should also be extracted and left with just data that a image or audio existed there at one point.}''
The next finding highlights user opinions regarding how (when) this data extraction should occur. 

\textbf{Data Should Be Extracted and Anonymized Before It Reaches the Researchers.}
Users preferred that the data be filtered down to the desired metadata even before the researchers could access it.
We informed the survey participants to understand \textit{data extraction} as the process of filtering the WhatsApp chat data down to only what is needed serving both as saving resources and protecting privacy by not collecting extraneous data. 
52.7\% of participants reported that an app should be used to perform the WhatsApp data extraction before it is sent to the researchers (30.6\% disagreed and 16.8\% neutral). In contrast, 21.0\% reported the researchers should conduct the WhatsApp data extraction, either on their own or with servers (59.9\% disagreed and 19.2\% neutral). 
For instance, S328 shared:
    ``\textit{I do believe their should be a system in place that extracts any unneeded data before being gathered}.''
    
Participants preferred that the data also be anonymized before researcher access. That is, 92.6\% reported (4.5\% neutral and 3\% disagreed) that personal identifiers in WhatsApp chats should be anonymized prior to researcher access versus the 70.1\% that supported (12.3\% neutral and 17.7\% disagreed) researchers anonymizing the WhatsApp data themselves. S71 explained: ``\textit{I believe any identifying data should be completely removed. Having it removed automatically before researchers get it should be the way to have it done so there is no potential to breach a person's information}.''

\textbf{Participants Should See (And Possibly Edit) The Data Before It Is Sent to Researchers.}
Almost every participant reported (92.9\% agreed,  3\% disagreed, and 4.2\% were neutral) that they should be able to view the data being collected by the researchers. However, participants had less agreement when it comes to the ability to edit extracted data. 52.1\% agreed and 24.6\% disagreed with allowing users to edit data before it is sent to the researchers (23.4\% were neutral), indicating a slight majority supporting editable data. S70 explains their support: ``\textit{the data owner should have full control over how much is collected and how it us used.}'' Unfortunately, no participant directly spoke to their disagreement with allowing data to be editable. A possible reason for their position may be that they think it could make the data less accurate or could be abused as a sort of `tampering.'

\subsubsection{Collecting Chat Data With a Locally Running Application Raises Concerns About Group Consent}
When asked about consent, survey respondents indicated that every user in the chat should be involved with the study when it comes to informed consent, not just a sole study participant in contact with the research team and operating on behalf of all chat members.
This indicates that the UCDS principles should be expanded further to meet user preferences related to researcher-to-group communication.

\textbf{Informed Consent Required From Every Chat Member.}
Almost every participant (96.1\%) agreed that even if only one chat member was directly involved in a research study, they should receive consent from the other group members before any data is collected. Only 1.5\% disagreed, and the remaining 2.4\% were neutral. Even if the data were to be sufficiently anonymized and could not be linked to any chat members, most participants still wanted all members to be informed and provide consent, as S216 said: ``\textit{The chat member should not send anyone else's data without their permission.}'' S264 elaborated that if other members' data is being used, then they should be treated like a normal study participant:
    ``\textit{It seems odd that the rest of the contacts in the group aren't being given agreement and disclosure forms that are usually given to each participant in a study.}''

\textbf{Users May Not Trust Their Fellow Chat Members.}
Participants were split on whether they trusted fellow chat members and their WhatsApp contacts to appropriately share data with researchers. 41.0\% of participants agreed that they trusted their WhatsApp contacts versus 37.2\% that disagreed (21.9\% were neutral).
Being in a chat together did not mean users would allow one another to share data on their behalf. S319 said that ``\textit{This feels like a huge violation of privacy and trust. If I found out that somebody I was writing to was doing this, I wouldn't write to them again. They can't be trusted.}''

\subsection{Study Part 2 Limitations}
Study Part~2 had several limitations. While, crowd-sourced data collected from Prolific may include biases such as the \textit{rapid-responder bias}~\cite{prolific_bias}, researchers found Prolific data to be of higher quality than from popular competitors like Mechanical Turk~\cite{peer_beyond_2017} and even more representative of the population than a census-representative web-panel~\cite{redmiles_how_2019}. We iterated our survey wording refining clarity based on pilot feedback, included a mix of open-ended and closed-ended responses, and randomized the order of survey statements where appropriate. Yet, the true preferences of survey participants may differ from their responses due to survey-related imperfections such as response errors, mode effects, or lack of understanding of the research scenarios. Furthermore, the participants were limited to WhatsApp users in the US only, and WhatsApp usage and perceptions could differ by country.

\section{Discussion} \label{Discussion}
Study Part~1 outlined proposed UCDS principles, provided an example implementation of them, and proved the feasibility of deploying the principles for research purposes.
Moreover, Study Part~2's survey suggested that the standard methods of joining and scraping public WhatsApp chats miss a majority of WhatsApp chats (private chats) and violate privacy expectations since users do not expect researchers to join their chats and read their contents. 
Our survey participants also preferred WhatsApp data be restricted to just necessary metadata, anonymized, and given participant oversight before researchers could access this data, in line with UCDS principles. 
The survey also raised concerns about the complications of researcher-to-group communication in WhatsApp data collection. Below, we discuss the implications of our findings, how UCDS fits into the broader picture of data collection, and how future work can improve and expand on the UCDS principles to shift data collection towards a two-sided data sharing relationship between users and researchers. 

\textbf{Researcher-to-Group Interface}
In developing URL-EXTRACTOR-APP as an example of UCDS, we surmised that following consenting procedures for a group member alongside increased data sanitization and privacy considerations would address user concerns about a fellow chat member sharing data from their chats.
However, participants made clear in Study Part~2 that interfacing with a sole chat member was not enough, despite that common current methods of collecting WhatsApp data do not typically ask for consent from \textit{any} chat member. That is, all chat members should be involved in the research process \textit{in addition to} the increased privacy considerations of UCDS, indicating that additional principles (e.g.,\ Group Engagement) may be necessary. Involving each chat member may also allow for addressing problematic cases where members of the same chat may not trust one another to appropriately share data with researchers.
Obtaining group consent is nontrivial and the frameworks for doing so are open questions in other fields~\cite{shaw2008ethics, schrag2006research, grill2009liberalism}. 
The extent to which each chat member should be involved and the mechanism for researcher-to-group communication could be explored in future work. 
Future work that adapts and tests the UCDS principles in other scenarios---not WhatsApp chat data---would also be invaluable for expanding and iterating the UCDS principles. 

\textbf{Expanding USDS: A Family of Related Methods}
The proposed UCDS principles from this work represent a family of related methods for collecting data. An app like URL-EXTRACTOR-APP is one potential instance of the principles focusing on URL sharing in WhatsApp chats. In general, the UCDS principles can be implemented in a number of study-specific ways or used alongside other privacy preserving methods. For example, implementations could be configured to target different types of metadata as opposed to URLs or could be paired with interviews to gain both a deeper understanding of a target population and privacy-preserving insights into their chats. Researchers could also set up a website to continually receive data that an app like ours locally collected from participants' devices, similar to a crowd-sourced tip line model~\cite{kazemi_tiplines_2021, tardaguila_whats_2020, whatscrap_whats_nodate, schwind2018whatsanalyzer} but with localized processing and metadata across a full chat instead of researcher controlled processing and stand-alone messages. If metadata is not sufficient for a research study's goals, researchers could tweak the collected metadata towards their research goals and still follow the design considerations. For example, in misinformation research, a study could include deploying an app that locally analyzes chats by perceptual hashing~\cite{kulshrestha_identifying_2021}. If actual chat content is required for analysis, researchers could implement the analysis locally where users can oversee and verify its results, expanding on what we did with our example app.

\textbf{Towards Best Practices In Data Collection}
Study Part~2's findings suggest that the discussion regarding the collection of public data should extend beyond studying WhatsApp for research purposes. Our participants agreed that just because their chat data may be technically accessible on the web, does not mean that they feel comfortable with researchers collecting it.
The decades-old Belmont Report serves as a wide-reaching foundation of ethical research standards for many disciplines~\cite{united1978belmont}. However, researchers have noted how the reports principles are not sufficient for modern online research and argue for additional ethical considerations for general online data collection. Further, while regulations are intended for platforms, their stipulations resemble the preferences of our study's participants. For instance, modern privacy regulations like the California Consumer Privacy Act and its extension, the California Privacy Rights Act, grant expansive rights to users over their data. These rights include knowing when your data is being collected, knowing how it is being used, and more control over collected data, all of which are supported by UCDS. The research community could also benefit from adapting these publicly supported rights to maintain the public's trust.

\section{Conclusion}
To investigate ethical data collection practices on WhatsApp, we conducted a two-part study. In Study Part~1, we outlined four UCDS principles for collecting WhatsApp chat data granting users control and transparency in data sharing processes while also minimizing data collected. We provided a proof-of-concept evaluation of following UCDS principles by creating and deploying an app using UCDS to build an exploratory dataset that led to valuable insights into participants' WhatsApp chats.
In Study Part~2, we surveyed WhatsApp users for their perceptions about data collection research methods including UCDS. Our findings suggest that users preferred how the UCDS principles handle data collection but highlighted that these principles could be expanded by exploring mechanisms for group consent. Future work can expand on the UCDS principles and evaluate them in other contexts to help the usable privacy community to continue to build best practices in data sharing between users and researchers.

\bibliographystyle{plain}
\bibliography{refs}

\appendix
\section{Supplementary Tables}
\label{supmat}

Here we include additional tables for this work.

\subsection{Dataset of Shared Chats}
\label{sec:extra:table}

\begin{table}[H]
    \centering
    \resizebox{0.8\columnwidth}{!}{%
    \begin{tabular}{|l|c|r|}
    \hline
\textbf{Participant ID} & \textbf{Chat ID} & \multicolumn{1}{c|}{\textbf{\# Members}} \\ \hline
P1                   & A             & 2                                         \\ \hline
P1                   & B             & 2                                         \\ \hline
P1                   & C             & 2                                         \\ \hline
P2                   & A             & 2                                         \\ \hline
P2                   & B             & 2                                         \\ \hline
P2                   & C             & 2                                         \\ \hline
P2                   & D             & 2                                         \\ \hline
P2                   & E             & 2                                         \\ \hline
P3                   & A             & 2                                         \\ \hline
P3                   & B             & 2                                         \\ \hline
P3                   & C             & 2                                         \\ \hline
P3                   & D             & 2                                         \\ \hline
P3                   & E             & 2                                         \\ \hline
P4                   & A             & 6                                         \\ \hline
P4                   & B             & 2                                         \\ \hline
P4                   & C             & 2                                         \\ \hline
P4                   & D             & 2                                         \\ \hline
P4                   & E             & 2                                         \\ \hline
P5                   & A             & 8                                         \\ \hline
P5                   & B             & 2                                         \\ \hline
P5                   & C             & 2                                         \\ \hline
P6                   & A             & 2                                         \\ \hline
P6                   & B             & 2                                         \\ \hline
P6                   & C             & 2                                         \\ \hline
P6                   & D             & 2                                         \\ \hline
P7                   & A             & 25                                        \\ \hline
P8                   & A             & 2                                         \\ \hline
P8                   & B             & 2                                         \\ \hline
P8                   & C             & 2                                         \\ \hline
P9                   & A             & 4                                         \\ \hline
P9                   & B             & 3                                         \\ \hline
P9                   & C             & 5                                         \\ \hline
P9                   & D             & 2                                         \\ \hline
P10                  & A             & 2                                         \\ \hline
P10                  & B             & 2                                         \\ \hline
P10                  & C             & 2                                         \\ \hline
    \multicolumn{2}{|r|}{median:}        & \multicolumn{1}{r|}{2}                    \\ \hline
    \end{tabular}%
    }
    \caption{The number of members for all chats in the dataset.}
    \label{tab:chat_size}
\end{table}

\hfill
\clearpage
\section{Survey Instruments}

Here we include the survey instruments for both the data-sharing participants of Study Part~1 and the survey participants of Study Part~2.

\subsection{Study Part~1: Background Survey Instrument}
\label{bsi}
In this survey we asked participants for their demographic information and characteristics of their WhatsApp usage.

\subsubsection{Demographics}\hfill\\

\textbf{First, we have a few background questions.}

\paragraph{1. }What is your age?
\begin{itemize}

\item	18-24 years old 
\item	25-34 years old 
\item	35-44 years old 
\item	45-54 years old 
\item	55 years or older 
\item	Prefer not to answer 
\end{itemize}

\paragraph{2. }What is your highest level of education?
\begin{itemize}

\item	Less than High School Degree 
\item	High School Degree 
\item	Associate’s/Some College Degree 
\item	Bachelor’s Degree 
\item	Master’s Degree 
\item	Doctoral Degree 
\item	Prefer not to answer 
\end{itemize}

\paragraph{3. }How long have you used WhatsApp?
\begin{itemize}
    \item Slider from 0 to 12 years
\end{itemize}

\paragraph{4. }What is estimated annual income? (If prefer not to answer, leave blank)
\begin{itemize}
    \item Drop-down menu
\end{itemize}

\paragraph{5. }What is your occupation? (If prefer not to answer, leave blank)
\begin{itemize}
    \item Text entry
\end{itemize}

\paragraph{6. }What state do you live in? (If prefer not to answer, leave blank)
\begin{itemize}
    \item Drop-down menu
\end{itemize}

\paragraph{7}What is your gender? 
\begin{itemize}
    \item Male
    \item Female
    \item Non-binary
    \item Other
\end{itemize}

\subsubsection{WhatsApp}\hfill\\

\textbf{Now, we'd like to know more about your WhatsApp usage. Please respond to the best of your ability.}

\paragraph{8}Do you consider yourself a frequent WhatsApp user?
\begin{itemize}
\item	Yes
\item No
\item Prefer not to say
\end{itemize}

\paragraph{9}Why do you use WhatsApp? (vs imessage, messenger, etc.)
\begin{itemize}
\item	More convenient for US communication
\item More convenient for international communication
\item Forced to join chat (work or family related)
\item Other (Please Specify):
\end{itemize}

\paragraph{10}How many contacts do you have on WhatsApp?
\begin{itemize}
\item Slider from 0 to 500
\end{itemize}

\paragraph{11}What relationship do you have with your WhatsApp contacts?
\begin{itemize}
\item Friends
\item Family
\item Work colleagues
\item Acquaintances
\item Prefer not to say
\end{itemize}

\paragraph{12a}What percentages of your WhatsApp messages are Groupchats (chats with more than 1 person)?
\begin{itemize}
\item slider from 0 to 100
\end{itemize}

\paragraph{12b}What percentages of your WhatsApp messages are Private Messages (chats with only 1 other person)?
\begin{itemize}
\item slider from 0 to 100
\end{itemize}

\paragraph{13}Are you aware of how WhatsApp encrypts your chats?
\begin{itemize}
\item Yes
\item No
\item Prefer not to say
\end{itemize}

\subsubsection{Misinformation Encounters and Recent Events}\hfill\\

\textbf{Next, we have some questions on your experience with sharing information in WhatsApp.}

\paragraph{14}Do you frequently rely on WhatsApp for news?
\begin{itemize}
\item Yes
\item No
\item Prefer not to say
\end{itemize}

\paragraph{15}How often do you \textbf{SEND} WhatsApp messages?
\begin{itemize}
\item Every day
\item A few times a week
\item Once a week
\item A few times a month
\item Rarely
\end{itemize}

\paragraph{16}How often do you \textbf{RECEIVE} WhatsApp messages?
\begin{itemize}
\item Every day
\item A few times a week
\item Once a week
\item A few times a month
\item Rarely
\end{itemize}

\paragraph{17}How often do you \textbf{CHECK} your WhatsApp messages?
\begin{itemize}
\item Every day
\item A few times a week
\item Once a week
\item A few times a month
\item Rarely
\end{itemize}

\paragraph{18}How often do you \textbf{FORWARD links/URLS} you receive in your WhatsApp chats?
\begin{itemize}
\item Every day
\item A few times a week
\item Once a week
\item A few times a month
\item Rarely
\end{itemize}

\paragraph{19}How often do you \textbf{FORWARD audio/video} messages you receive in your WhatsApp chats?
\begin{itemize}
\item Every day
\item A few times a week
\item Once a week
\item A few times a month
\item Rarely
\end{itemize}

\paragraph{20}How often do you \textbf{FORWARD images} messages you receive in your WhatsApp chats?
\begin{itemize}
\item Every day
\item A few times a week
\item Once a week
\item A few times a month
\item Rarely
\end{itemize}

\paragraph{21}How often do you \textbf{RECEIVE links} on your WhatsApp chats?
\begin{itemize}
\item Every day
\item A few times a week
\item Once a week
\item A few times a month
\item Rarely
\end{itemize}

\paragraph{22}How often have you \textbf{RECEIVED} messages on your WhatsApp chats that were false or misleading?
\begin{itemize}
\item Every day
\item A few times a week
\item Once a week
\item A few times a month
\item Rarely
\end{itemize}

\paragraph{23}How did you know that a message you received on WhatsApp was false or misleading?
\begin{itemize}
\item Person who sent it
\item Search on the web (Google or other search engines)
\item Looked suspicious
\item Known source of misinformation
\item Other (Please specify)
\end{itemize}

\paragraph{24}What did you do to fact check WhatsApp messages you found false or misleading?
\begin{itemize}
\item Online fact checkers
\item Through friends
\item Just knew
\end{itemize}

\paragraph{25}Do you consider yourself a frequent Instagram user?
\begin{itemize}
\item Yes
\item No
\item Prefer not to say
\end{itemize}

\begin{figure}[H]
    \centering
    \includegraphics[width=\columnwidth]{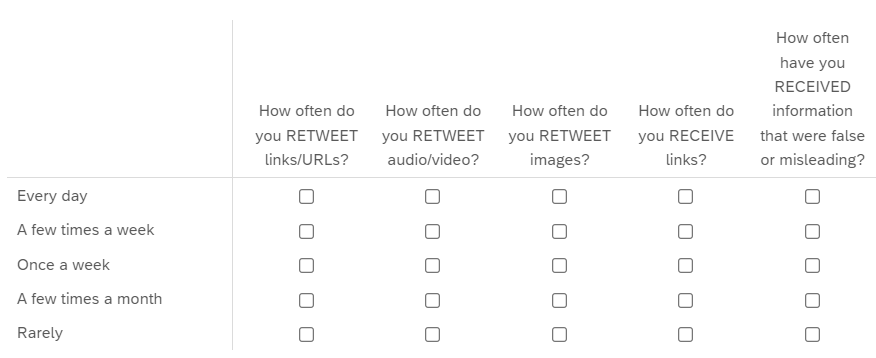}
    \caption{}
    \label{fig:survey_grid}
\end{figure}

\paragraph{26}Please answer the following on information you've seen \textbf{on Instagram} (Figure~\ref{fig:survey_grid}). \textit{Note: This question was only shown to participants that answered Yes to the previous question.}

\paragraph{27}Do you consider yourself a frequent Twitter user?
\begin{itemize}
\item Yes
\item No
\item Prefer not to say
\end{itemize}

\paragraph{28}Please answer the following on information you've seen \textbf{on Twitter} (Figure~\ref{fig:survey_grid}). \textit{Note: This question was only shown to participants that answered Yes to the previous question.}

\paragraph{29}Do you consider yourself a frequent Facebook/Messenger user?
\begin{itemize}
\item Yes
\item No
\item Prefer not to say
\end{itemize}

\paragraph{30}Please answer the following on information you've seen \textbf{on Facebook/Messenger} (Figure~\ref{fig:survey_grid}). \textit{Note: This question was only shown to participants that answered Yes to the previous question.}

\paragraph{31}Please answer the following on information you've seen \textbf{on your Text messages} (Figure~\ref{fig:survey_grid}). 

\subsubsection{Technology}\hfill\\

\textbf{Next, we have some questions on your thoughts with any current technology.}

\paragraph{32}Have you used any of these features on WhatsApp?
\begin{itemize}
\item WHO Health Alert (screenshot provided)
\item Magnifying glass icon pop up (screenshot provided)
\item No
\item Prefer not to say
\end{itemize}

\paragraph{33}Would you want a better way to fact-check information on WhatsApp?
\begin{itemize}
\item Yes
\item No
\item Prefer not to say
\end{itemize}

\paragraph{34}Would you want WhatsApp to cover or censor known misinformation sources?
\begin{itemize}
\item Yes
\item No
\item Prefer not to say
\end{itemize}

\paragraph{35}In what ways, do you think WhatsApp can be improved to help address these issues with false, inaccurate, or misleading information?
\begin{itemize}
    \item Open text entry. 
\end{itemize}

\subsubsection{Conclusion}\hfill\\

\textbf{Please include any final thoughts you would like to share below.}

\paragraph{36}How has anything you said been vastly different from how you send or receive messages on other social media platforms (Instagram, Twitter, Facebook, etc.) you use?
\begin{itemize}
    \item Open text entry. 
\end{itemize}

\paragraph{37}Is there anything else regarding WhatsApp, tech, or information quality that you want to talk about?
\begin{itemize}
    \item Open text entry. 
\end{itemize}

\subsection{Study Part~2: User Perceptions Survey Instrument}
\label{user_perceptions_survey}
In this survey we asked participants for characteristics of their WhatsApp usage, their perceptions of two data collection scenarios, and their comfort sharing different types of chat data with researchers. We concluded the survey with miscellaneous data collection questions and optional demographic questions. 

\subsubsection{WhatsApp Usage}\hfill\\

\textbf{Now, we'd like to know more about your WhatsApp usage. Please respond to the best of your ability.}

\paragraph{1}Do you consider yourself a frequent WhatsApp user?
\begin{itemize}
    \item Yes
    \item No
\end{itemize}

\paragraph{2}How long have you used WhatsApp?
\begin{itemize}
    \item Slider from 0 to 13 years
\end{itemize}

\paragraph{3}Why do you use WhatsApp (instead of other messaging apps like iMessage, Facebook Messenger, Signal, etc.)? Select all that apply. 
\begin{itemize}
    \item More convenient for US communication
    \item More convenient for International communication
    \item My family and friends use it
    \item My work uses it
    \item Works better than the other services
    \item Other (Please Specify):
\end{itemize}

\paragraph{4}Please select only WhatsApp from the list below. \textit{Attention Check Question.}
\begin{itemize}
    \item WhatsApp
    \item Telegram
    \item Signal
    \item Facebook Messenger
    \item Threema
    \item Wire
    \item Other
\end{itemize}

\paragraph{5}How many contacts do you have on WhatsApp?
\begin{itemize}
    \item Slider from 0 to 500 with an option to check a box indicating over 500 contacts.
\end{itemize}

\paragraph{6}What relationship do you have with the members of your WhatsApp chats? Select all that apply.
\begin{itemize}
    \item Friends
    \item Family
    \item Work colleagues
    \item Acquaintances
    \item People I know only from WhatsApp group chats
    \item Other (Please Specify):
\end{itemize}

\paragraph{7}How have you joined chats in the past? Select all that apply.
\begin{itemize}
    \item Created my own chats with my WhatsApp contacts
    \item My friends/family created chats and added me
    \item By receiving invite links from friends/family
    \item By finding invite links online from public WhatsApp groups
    \item Other (Please Specify):
\end{itemize}

\subsubsection{Scenario Intro}\hfill\\

\textbf{In these questions, we want to know what methods you feel are okay for researchers to use for the collection of WhatsApp chat data. You will be given 2 scenarios, each with a page worth of questions. Please answer each full page of questions based on the scenario given at the top.} \textit{Note that Scenario 1 and 2 were presented in random order.}

\subsubsection{Data Collection Method Scenario 1}\hfill\\

Imagine the following scenario: \textbf{The invite link to one of your WhatsApp chats is publicly available online, and a researcher uses it to join the chat.} Please answer all the questions on this page based on this scenario.

\begin{figure}[H]
    \centering
    \includegraphics[width=\columnwidth]{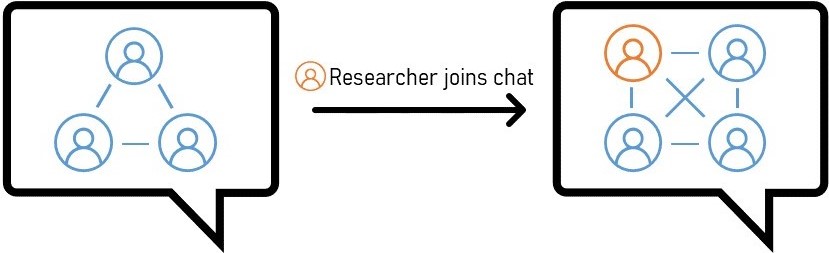}
\end{figure}

\paragraph{8}What should the researcher do when they first join the chat? \textit{5-point Likert (Strongly Disagree to Strongly Agree)}
\begin{itemize}
    \item The researcher should inform the chat members of their presence.
    \item The researcher should inform the chat members of their intention to collect data.
    \item The researcher should carry on as a normal chat member.
    \item The researcher does not need to say anything.
\end{itemize}

\paragraph{9}Should the researcher ask for permission to collect data? \textit{5-point Likert (Strongly Disagree to Strongly Agree)}
\begin{itemize}
    \item The researcher should obtain consent from all chat members.
    \item The researcher should obtain consent from a majority of chat members but not necessarily all of them.
    \item The researcher should obtain consent from at least one chat member but not necessarily a majority of them.
    \item The researcher need not worry about consent because the chat is publicly accessible.
    \item The researchers can collect chat data without consent as long as they provide the ability for chat members to opt-out.
    \item The researcher should only collect data from the members that have provided consent.
    \item If consent is needed, the researchers should obtain it by directly messaging the individual chat members.
    \item If consent is needed, the researchers should obtain it by messaging to the whole group chat.
\end{itemize}

\paragraph{10}Should the researcher collect personally identifiable information (e.g.,\ Names, Profile Pictures, Phone Numbers) of the chat members? \textit{5-point Likert (Strongly Disagree to Strongly Agree)}
\begin{itemize}
    \item It is okay if the researcher collects any personally identifiable information of the chat members.
    \item The researchers should not collect any personally identifiable information of the chat members.
\end{itemize}

\paragraph{11}What else would you like to say about this data collection scenario?
\begin{itemize}
    \item Open text entry.
\end{itemize}

\subsubsection{Data Collection Method Scenario 2}\hfill\\

Imagine the following scenario: \textbf{One of your WhatsApp contacts is involved in a research study that is collecting chat data, including data from a chat that you are also in. In this scenario, the researchers do not join your chat but are in contact with one of the chat members.} Please answer all the questions on this page based on this scenario.

\begin{figure}[H]
    \centering
    \includegraphics[width=\columnwidth]{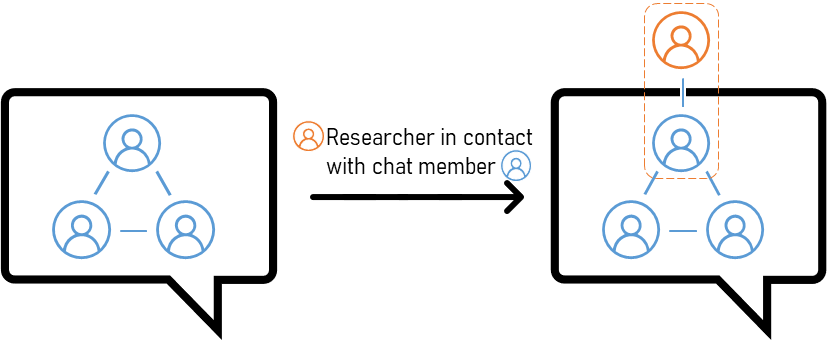}
\end{figure}

\paragraph{12}What should your WhatsApp contact do? \textit{5-point Likert (Strongly Disagree to Strongly Agree)}
\begin{itemize}
    \item They should inform you of their involvement in the study.
    \item They should ask for your permission to share chat data for the chats you are a member of.
    \item They can continue with the study without any additional considerations.
    \item They can continue with the study without talking with you as long as no data can be linked to you.
\end{itemize}

Here are some examples of WhatsApp chat data.

\begin{figure}[H]
    \centering
    \includegraphics[width=\columnwidth]{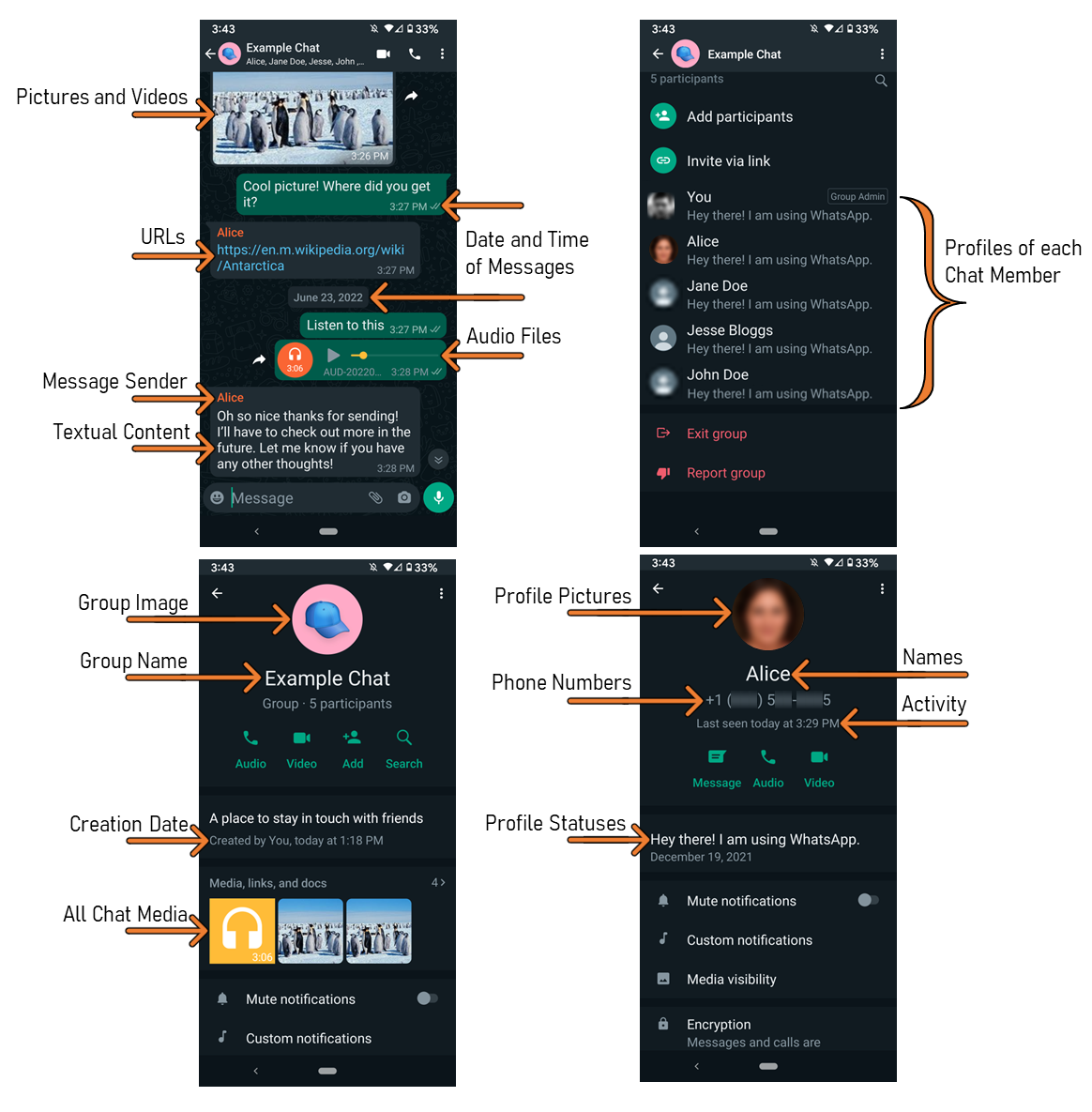}
\end{figure}

\paragraph{13}As shown above, a single WhatsApp chat contains a lot of different data. Researchers often don't need all of it, so they extract only the pieces that they may need for their study. The statements below are about how this data extraction occurs. \textit{5-point Likert (Strongly Disagree to Strongly Agree)}
\begin{itemize}
    \item Your WhatsApp contact should upload the full chat data to the researcher's online servers that will automatically perform the data extraction.
    \item Your WhatsApp contact should send the full chat data directly to the researchers who will manually perform the data extraction.
    \item Your WhatsApp contact should use an app that automatically performs the data extraction before sending it to the researchers.
\end{itemize}

\paragraph{14}WhatsApp chat data includes personal identifiers, such as Names or Phone Numbers. Since researchers typically do not need this information, they anonymize the data by replacing personal identifiers with placeholders like User0 and User 1. The statements below are about how the anonymization occurs. \textit{5-point Likert (Strongly Disagree to Strongly Agree)}
\begin{itemize}
    \item The personal identifiers do not need to be anonymized.
    \item The personal identifiers should be anonymized by the researchers.
    \item The personal identifiers should be automatically anonymized before the researchers get it.
\end{itemize}

\paragraph{15}What else would you like to say about this data collection scenario?
\begin{itemize}
    \item Open text entry.
\end{itemize}

\subsubsection{Data Sharing Comfort by Type}\hfill\\

\textbf{In these questions, we want to know what parts of your chat data you feel comfortable sharing with researchers.}

\paragraph{16}Please indicate the level of comfort you feel towards sharing each category of your chat data with researchers. \textit{5-point Likert (Extremely Uncomfortable to Extremely Comfortable). The statements below were presented in a random order while keeping similar categories next to each other to reduce cognitive burden on the participants (e.g.,\ Whether a Message Contains a URL and The URL Itself would always be adjacent).}
\begin{itemize}
    \item The number of users in the chat
    \item The names of all chat members
    \item The contact information (e.g.,\ phone numbers) of the chat members
    \item The anonymized sender of each message (e.g.,\ Jane Doe and John Doe become User-0 and User-1, respectively)
    \item Whether each message contains a URL
    \item The URL itself
    \item Whether each message contains text
    \item The full textual content
    \item Whether a message contains media (audio, video, or image)
    \item The media itself (the audio, video, or media)
    \item To make sure you are reading the questions carefully, please select Somewhat comfortable for this statement. \textit{Attention Check Question}
    \item Whether a message contains a document attachment
    \item The document itself
    \item The start and end dates of a chat
    \item The calendar date each message was sent
    \item The time of day each message was sent
    \item The number of messages from each user per day
    \item The total number of messages from each user
    \item The total number of messages from each user by message type (e.g.,\ text, media, URL, document)
    \item Any and all data related to the chat
\end{itemize}

\subsubsection{Miscellaneous Data Collection Questions}\hfill\\

\paragraph{17}What controls should you have over the chat data being collected? \textit{5-point Likert (Strongly Disagree to Strongly Agree)}
\begin{itemize}
    \item I should be able to see the data that the researchers collect.
    \item I should be able to edit the data before it is sent to the researchers.
\end{itemize}

\paragraph{18}How do you feel about the roles of others in the collection of your WhatsApp chat data? \textit{5-point Likert (Strongly Disagree to Strongly Agree)}
\begin{itemize}
    \item I trust researchers to be responsible with the data they collect from my WhatsApp chats.
    \item I trust my WhatsApp contacts to appropriately share chat data with researchers.
    \item I trust myself to appropriately share chat data with researchers.
\end{itemize}

\paragraph{19}How do you feel about your chat data being used for research? \textit{5-point Likert (Strongly Disagree to Strongly Agree)}
\begin{itemize}
    \item I should be compensated for my WhatsApp chat data.
    \item I should be informed if my WhatsApp chat data is being collected.
    \item I should be informed how my WhatsApp chat data is being used.
    \item I would like it if my chat data is being used for research purposes.
    \item My willingness to share my WhatsApp data with researchers depends on the goals of the research.
\end{itemize}

\paragraph{20}Is there anything else you would like to add about WhatsApp data collection?
\begin{itemize}
    \item Open text entry.
\end{itemize}

\subsubsection{Demographics}\hfill\\

\paragraph{21}What is your age?
\begin{itemize}
\item	18-24 years old 
\item	25-34 years old 
\item	35-44 years old 
\item	45-54 years old 
\item	55 years or older 
\item	Prefer not to answer 
\end{itemize}

\paragraph{22}What is the highest level of school you have completed or the highest degree you have received? 
\begin{itemize}
    \item Less than high school degree
    \item High school graduate (high school diploma or equivalent including GED)
    \item Some college but no degree
    \item Associate degree in college (2-year)
    \item Bachelor's degree in college (4-year)
    \item Master's degree
    \item Doctoral degree
    \item Professional degree (JD, MD)
    \item Prefer not to answer
\end{itemize}

\paragraph{23}What is estimated annual income?
\begin{itemize}
    \item Less than \$25,000
    \item \$25,000 to \$50,000
    \item \$50,000 to \$100,000
    \item \$100,000 to \$200,000
    \item More than \$200,000
    \item Prefer not to say
\end{itemize}

\paragraph{24}What state do you live in?
\begin{itemize}
    \item Drop-down menu including option for Prefer not to say.
\end{itemize}

\paragraph{25}What is your gender? 
\begin{itemize}
    \item Female identifying
    \item Male identifying
    \item Non-binary
    \item Prefer not to answer
    \item Prefer to self-describe:
    \begin{itemize}
        \item text entry
    \end{itemize}
\end{itemize}

%%%%%%%%%%%%%%%%%%%%%%%%%%%%%%%%%%%%%%%%%%%%%%%%%%%%%%%%%%%%%%%%%%%%%%%%%%%%%%%%
\end{document}